\RequirePackage{fix-cm}
\documentclass{svjour3}                     
\AtBeginDocument{ \paperwidth=\dimexpr 1in + \oddsidemargin + \textwidth + 1in + \oddsidemargin \relax \paperheight=\dimexpr 1in + \topmargin + \headheight + \headsep + \textheight + 1in 
+ \topmargin \relax \usepackage[pass]{geometry} \relax }
\usepackage[T1]{fontenc} 
\usepackage{tgtermes, tgheros} 
\usepackage{amsfonts}
\usepackage{amsmath}

\usepackage{marvosym}
\usepackage{amsthm}
\usepackage{mathtools}
\usepackage{optidef}
\usepackage{stmaryrd}
\usepackage{graphicx, color}
\usepackage{hyperref}
\usepackage{natbib}
\usepackage{caption}
\usepackage{subcaption}
\usepackage{layout}
\usepackage{comment}
\usepackage{setspace}
\usepackage{lscape}
\usepackage{dcolumn}
\usepackage{booktabs}
\usepackage{multirow}
\usepackage{threeparttable}
\usepackage{tabularx}
\usepackage[table]{xcolor}
\newtheorem{propo}{Proposition}
\journalname{ }
\def\makeheadbox{\relax}
\begin{document}
\title{
Optimal longevity of a dynasty
} 
\titlerunning{ }        
\author{Satoshi Nakano \and Kazuhiko Nishimura}
\institute{
Satoshi Nakano \\ 0000-0003-0424-585X 
 \at Nihon Fukushi University, Tokai, 477-0031, Japan \\
\email{nakano@n-fukushi.ac.jp}
\and 
Kazuhiko Nishimura (\Letter) \\ 0000-0002-7400-7227 
 \at Chukyo University, Nagoya, 466-8666, Japan \\
\email{nishimura@lets.chukyo-u.ac.jp}
}
\date{\today}

\maketitle
\begin{abstract}
Standard optimal growth models implicitly impose a ``perpetual existence'' constraint, which can ethically justify infinite misery in stagnant economies. This paper investigates the optimal longevity of a dynasty within a Critical-Level Utilitarian (CLU) framework. By treating the planning horizon as an endogenous choice variable, we establish a structural isomorphism between static population ethics and dynamic growth theory. Our analysis derives closed-form solutions for optimal consumption and longevity in a roundabout production economy. We show that under low productivity, a finite horizon is structurally optimal to avoid the creation of lives not worth living. This result suggests that the termination of a dynasty can be interpreted not as a failure of sustainability, but as an \textit{altruistic termination} to prevent intergenerational suffering. We also highlight an ethical asymmetry: while a finite horizon is optimal for declining economies, growing economies under intergenerational equity demand the ultimate sacrifice from the current generation.
\keywords{Critical-level Utilitarianism \and Dynamic Programming \and Optimal Population \and Intergenerational Equity}
\JEL{O21 \and I30 \and C61}
\end{abstract}

\clearpage
\section{Introduction}

How long should a dynasty last, if each generation's welfare must meet a morally acceptable standard? Conventional economic models of optimal growth typically assume a perpetual planning horizon, justified by the mathematical convenience of asymptotic stability or the ethical stance of treating all future generations impartially \citep{Ramsey1928}. While this framework functions well in growing economies, it raises deep ethical concerns when applied to stagnant or declining economies with resource constraints.

\begin{figure}[t!]
\centering
\includegraphics[width=0.85\textwidth]{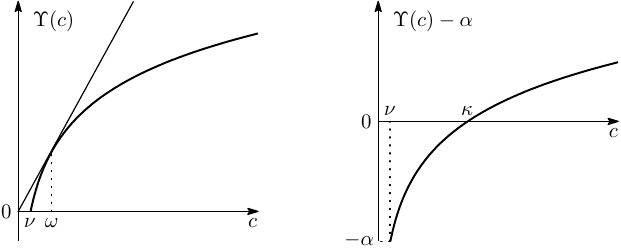}
\caption{
Left: Utility function of an individual. The consumption level for well-being subsistence is denoted by $\nu$.
Right: Contribution to the critical-level utilitarian population value, where $\alpha$ denotes the critical level. At $c=\kappa$, the individual utility reaches the critical level. While our logarithmic specification implies a mathematical lower bound $\nu > 0$, the structural mechanism preventing the Repugnant Conclusion in our model is driven strictly by the social threshold $\kappa$, not the biological subsistence $\nu$.
}
\label{Fig1}
\end{figure}

To illustrate this, consider a static utilitarian axiology where a social planner aims to find the population size $n$ that maximizes the population value $\mathcal{V}=n \Upsilon(c)$ under a fixed resource constraint (see, Figure \ref{Fig1} left). If the well-being subsistence level approaches zero ($\nu \to 0$), which \citet{CORDOBA} refers to as the homothetic case, classical utilitarianism concludes that the optimal consumption also approaches zero ($\omega \to 0$) while the optimal population size approaches infinity ($n \to \infty$). This leads to the Repugnant Conclusion \citep{PARFIT}, where a massive population living at the absolute margin of subsistence is ethically preferred to a smaller, prosperous one. In a dynamic macroeconomic setting, standard models implicitly impose a ``perpetual existence'' constraint \citep{DASGUPTA2019}. By enforcing an infinite horizon ($N \to \infty$) under homothetic preferences, these models may inadvertently justify a dynamic analogue of the Repugnant Conclusion: an infinite accumulation of sub-optimal lives.

To avoid such implications, this paper revisits dynastic planning through the lens of Critical-Level Utilitarianism (CLU) \citep{BlackorbyDonaldson1984, BBD2005}. Unlike standard utilitarianism, CLU introduces a normative threshold of well-being, or critical level ($\alpha > 0$), establishing a social threshold ($\kappa$) for a life worth living (see, Figure \ref{Fig1} right). 
An individual's existence contributes positively to social welfare if and only if their utility exceeds this critical level; otherwise, their addition is considered a negative contribution to the population value. 
Introducing $\alpha > 0$ structurally prevents the Repugnant Conclusion even if the underlying individual utility function is homothetic ($\nu \to 0$).

By applying this ethical principle to a dynamic setting, we treat the horizontal length of the dynasty---the number of generations $N$ (longevity)---as an endogenous choice variable.\footnote{This approach complements the dynamic optimal population literature, which has traditionally focused on endogenizing the vertical size of cohorts (i.e., the population growth rate) over an infinite planning horizon. This tradition was initiated by \citet{Samuelson1975} and extended to CLU frameworks by \citet{RS2011, RS2019}. By inverting this paradigm, our framework reveals a structural isomorphism: the Repugnant Conclusion (a massive population with sub-critical quality of life) and the dynamic accumulation of negative social value (an infinite dynasty with sub-critical quality of life) are two sides of the same coin.} If economic constraints, such as low productivity, force future consumption below the social threshold $\kappa$, the social evaluation of those generations becomes negative, even if their individual utility remains technically positive. Drawing on the population ethics of \citet{Fleurbaey2014}, who posit that when welfare falls below a neutral threshold, non-existence is strictly preferred, we argue that the termination of a dynasty is not a failure of sustainability, but an \emph{altruistic termination} to prevent the infinite accumulation of negative social value. Mathematically, this termination can be formulated as an irreversible optimal stopping problem, utilizing the two-stage optimal control framework \citep{Tomiyama1985} commonly applied in macroeconomic regime-switching models \citep[e.g.,][]{Boucekkine2013}.

We acknowledge that other axiomatic frameworks, such as Rank-Discounted Utilitarianism (RDU), have been proposed to avoid the Repugnant Conclusion \citep{AsheimZuber2014}. RDU prevents the conclusion by strictly discounting the value of additional lives. However, this approach can lead to the opposite extreme---an ``anti-population bias''---where a small population with high utility is preferred over a larger population that is also well-off. Furthermore, in a dynamic context, RDU implies that the welfare weight of future generations depends on their relative rank in history. This dependency destroys the recursive structure of the optimization problem and leads to time inconsistency. In contrast, CLU provides a transparent, absolute threshold for ``lives worth living'' that preserves both intergenerational equity and the tractability of dynamic programming.

The remainder of the paper proceeds as follows. Section~2 describes the model of roundabout production using Cobb--Douglas technology and defines the main problem as finite horizon dynamic programming. We analytically derive the optimal consumption path and the optimal longevity $N^*$. In Section~3, we analyze the properties of the solution under two parameter settings: the AK setting (unity output elasticity of capital) and the Zero Discounting (ZD) setting (intergenerational equity). We show that under specific conditions, a finite horizon is structurally optimal. Finally, to prove that our finite-horizon result is a robust structural consequence of CLU rather than an artefact of the logarithmic utility assumption, Appendix \ref{apd4} develops a continuous-time optimal stopping formulation with a general Constant Relative Risk Aversion (CRRA) utility function. Section~4 concludes with discussions on the ethical implications of our findings.

\section{Model}
\subsection{Critical-Level Preferences}

Before describing the production technology, we first specify the social planner's objective function. 
We posit a social planner whose objective is to maximize the population value based on CLU. 
Let $\Upsilon(c)$ denote the lifetime utility (well-being) of an individual consuming a constant level of resources $c$.
The planner aims to determine the consumption path and the planning horizon (number of generations) to maximize the following social welfare function:
\begin{equation}
\mathcal{V} = \sum_{t=0}^{N} \left( \Upsilon(c_t) - \alpha \right)
\end{equation}
where $\alpha$ represents the critical level of utility.
An individual's life contributes positively to social welfare if and only if their utility exceeds this critical threshold $\alpha$;
otherwise, it reduces social welfare.

For analytical tractability and to capture the essential trade-offs, we employ a logarithmic utility function of the form $\Upsilon(c) = \log c + \alpha$.
This functional form is widely employed in optimal growth literature, including climate change economics \citep{Nordhaus1992, Stern2007}, which facilitates comparisons with standard benchmarks.
Under this specification, the contribution of generation $t$ to the population value becomes:
\begin{equation}
\Upsilon(c_t) - \alpha = (\log c_t + \alpha) - \alpha = \log c_t
\end{equation}
Consequently, the objective function simplifies to $\mathcal{V} = \sum_{t=0}^{N} \log c_t$.
We define the \emph{well-being subsistence level} $\nu$ as the consumption level where utility is exactly zero, i.e., $\Upsilon(\nu) = 0$.
In our logarithmic specification, this corresponds to $\nu = e^{-\alpha}$.
As illustrated in Figure~\ref{Fig1}, the domain of permissible consumption is $c > \nu$.
The critical level $\alpha$ thus serves as a ``buffer'' above the biological subsistence level.
The planner will only choose to extend the dynasty if the consumption level allows for a utility strictly greater than the critical level (i.e., $\log c_t > 0$, or $c_t > 1$ in normalized terms), ensuring that each generation enjoys a life worth living.

\subsection{Roundabout production}
Let us begin by postulating a two-factor Cobb--Douglas production function characterized by constant returns to scale, as follows:
\begin{align}
Y_t  = {A} (K_t)^\theta (L_t)^{1-\theta} 
\label{PF}
\end{align}
where $Y_t$ denotes the economy's output, $K_t$ denotes capital, and $L_t$ denotes labor, all of which are effective during period $t$.
For the relevant parameters, $0< \theta <1$ denotes the output elasticity of capital.
The level of technology is indicated by productivity, denoted by $A$, which is assumed to be fixed throughout the considered time span.
Note that ${B} \equiv {A} \theta^\theta (1-\theta)^{1-\theta}$ is the cost function-based productivity.\footnote{Otherwise, $B$ is referred to as the dual productivity.
See Appendix \ref{apd1} for more details.
}

The breakdown of output into investment, capital depreciation, and final consumption is described below:
\begin{align}
Y_t = K_{t+1} - K_{t} + \delta K_{t} + C_{t}
\label{balance}
\end{align}
Here, $\delta \leq 1$ denotes the capital depreciation factor, and $C_{t}$ denotes aggregate consumption in period $t$.
Combining equations (\ref{PF}) and (\ref{balance}) and dividing both sides by $L_t=L$, which we assume to be constant over time, leads to the following intertemporal dynamics (or roundaboutness) of capital intensity:
\begin{align}
y_t = {A} (k_t)^\theta = k_{t+1} - (1-\delta) k_{t} + c_{t}
\label{trans}
\end{align}
where $k_{t} = K_t/L$ and $k_{t+1}=K_{t+1}/L$ denote capital intensities in periods $t$ and $t+1$, respectively.
Additionally, $y_t = Y_t/L$ and $c_{t}=C_t/L$ denote per capita output and consumption, respectively, in period $t$.
The social planner aims to maximize the critical-level utilitarian population value $\mathcal{\mathcal{V}}$ as previously specified, subject to the state transition function (\ref{trans}), i.e.,
\begin{maxi!}
    {c_0, c_1, \cdots, c_N}          
    {\mathcal{\mathcal{V}}=\sum_{t=0}^{N} \beta^t \log c_t \label{obj}} 
    {\label{prb}}  
    {} 
    \addConstraint{k_{t+1}}{={A} (k_t)^\theta - c_t,  \label{const}}{\quad k_{N+1}=0,}
\end{maxi!}
given the initial state $k_0$,
we assume complete depreciation $\delta =1$ and introduce the discount factor $\beta \leq 1$.
We let $\beta \to 1$ for a (critical-level) utilitarian assessment.\footnote{
The discount factor $\beta$ is otherwise referred to as the rate of time preference.
In the context of dynastic optimization, $\beta$ may be referred to as the rate of generational preference (of the population).
}
The optimal consumption path $(c_{0}^*, c_{1}^*, \cdots, c_{N}^*)$ is clearly dependent on the planning horizon $N$.
We therefore solve the above problem hierarchically.
That is, we first solve for the optimal consumption path given $N$ via finite horizon dynamic programming and obtain the population value function with respect to the planning horizon $\mathcal{\mathcal{V}}[N]$;
therefore, we search for the optimal planning horizon $N^*$.

\subsection{Finite horizon dynamic programming}
The primary problem here is to solve for an optimal consumption path given a planning horizon.
The Bellman equation of the problem is as follows:\footnote{
Note that the objective function of the primary problem (\ref{obj}) is given at $t=0$, i.e., $\mathcal{V}[N] = \mathcal{V}_0 [k_0;
N]$.
}
\begin{align}
\mathcal{\mathcal{V}}_{t}\left[k_t ; \ N \right] = \max_{c_{t}} \left( \log c_{t} + \beta \mathcal{\mathcal{V}}_{t+1} \left[k_{t+1} = {A} (k_t)^\theta - c_t; \ N \right] \right)
\label{Bellman}
\end{align}
The optimal trajectory of the state variable $k_t$ is given by Lemma \ref{optkt}, which we append to Appendix \ref{apd2} with proofs, as follows:
\begin{align}
k_t^*[N] = (k_0)^{\theta^t} \prod_{i=1}^t \left( \frac{S_{{N}-i}}{S_{{N}-i+1}} {A} \beta \theta \right)^{\theta^{t-i}}
=
(k_0)^{\theta^t} \prod_{i=1}^t \left( \frac{S_{{N}-t+i-1}}{S_{{N}-t+i}} {A} \beta \theta \right)^{\theta^{i-1}}
\label{optktn}
\end{align}
where $S_{\ell}$ is defined as follows:
\begin{align}
S_\ell \equiv \sum_{i=0}^\ell (\beta \theta)^i 
=1 + (\beta \theta) + (\beta \theta)^2 + \cdots + (\beta \theta)^\ell
=\frac{1-(\beta\theta)^{\ell+1}}{1-\beta\theta}
\label{sum}
\end{align}
The following optimal trajectory of consumption is obtained by (\ref{optktn}) and (\ref{ctkt}) Appendix \ref{apd2}, which must be true according to 
the proof of Lemma \ref{p1}.
\begin{align}
c_{t}^*[N]
=
\frac{{A} \left( k_t^*[N] \right)^\theta}{S_{N-1}}
=
\frac{{A}(k_0)^{\theta^{t+1}}}{S_{{N}-t}} 
\prod_{i=1}^t \left( \frac{S_{{N}-t+i-1}}{S_{{N}-t+i}} 
{A} \beta \theta
\right)^{\theta^i}
\label{optct}
\end{align}
With formula (\ref{optct}), the population value function is given as follows:
\begin{align}
\mathcal{V}[N] &= \sum_{t=0}^N \beta^t \log c_{t}^*[N] \notag
\\
&=\log \left(\frac{{A} (k_0)^\theta}{S_{N}} \right)
+\sum_{t=1}^{N} \beta^t \log \left( 
\frac{{A}(k_0)^{\theta^{t+1}}}{S_{{N}-t}} 
\prod_{i=1}^t \left( \frac{S_{{N}-t+i-1}}{S_{{N}-t+i}} 
{A} \beta \theta
\right)^{\theta^i} \right)
\label{optval}
\end{align}
We hereafter aim to maximize this function with respect to the planning horizon $N$.
Our approach to analyzing the population value (\ref{optval}) adopts a numerical rather than an analytical framework because the derivative of $\mathcal{V}[N]$ with respect to $N$ does not seem to provide meaningful insights.
In the following section, the optimal trajectory of contributions $\log c_t^*[N]$ and the population value $\mathcal{V}[N]$, for any given planning horizon $N$, becomes manageable under $\theta = 1$ (known as the AK setting) and $\beta <1$.
We also find that the trajectory of contributions $\log c_t^*[\infty]$ and the population value $\mathcal{V}[\infty]$ for an infinite planning horizon is evaluable under $\beta \theta <1$.
We therefore base our study of ZD ($\beta =1$) on this parameter setting (i.e., $\beta \theta <1$, which indicates that $\theta <1$, consistent with a Cobb--Douglas model).
In the following section, we delve into the abovementioned two broad settings of parameters, namely, AK production with future discounting ($\theta=1$ and $\beta<1$), which we term AK settings, and Cobb--Douglas production without future discounting ($\theta<1$ and $\beta=1$), which we term ZD settings.
Table \ref{tab_parameters} summarizes the parameter settings chosen for the numerical examinations.
Note that cases I--IV correspond to AK settings whose solution paths are characterized by the sign of $\log (A\beta)$, whereas cases V--VII correspond to ZD settings whose solution paths are characterized by the sign of $\log B$.
Cases VIII and IX correspond to the parameter settings where $\log({A}\beta) = \log B = 0$.
\begin{table}[t]
\begin{center}
\begin{threeparttable}
\newcolumntype{.}{D{.}{.}{3}}
\newcolumntype{;}{D{.}{.}{1}}
\caption{Parameters applied in various cases.} \label{tab_parameters}
\begin{tabularx}{0.95\linewidth}{r  .  .  .  c  c  r  ;  ;
}
\hline\noalign{\smallskip}
 case & \multicolumn{1}{c}{${A}$} & \multicolumn{1}{c}{ $\beta$} & \multicolumn{1}{c}{$\theta$} & \multicolumn{1}{c}{ $\log ({A} \beta)$} & {$\log B$}&\multicolumn{1}{r}{$N^*$}  & \multicolumn{1}{r}{$\mathcal{V}[N^*]$}  & \multicolumn{1}{r}{$\mathcal{V}[\infty]$} \\
 \hline \noalign{\smallskip}
I & 1.012 & 0.992 & 1  & $+$ &  & $\infty$ & 84.7 & \\
II  & 1.01  & 0.992 & 1  & $+$ &  & $95$ & 60.0 &  \\
III  &  \multicolumn{1}{c}{$1/\beta$} & 0.992 & 1 & $0$ &  & $73$ & 55.6 & \\
IV  & 1.005  & 0.992 & 1 & $-$ &  & $58$ & 51.0 &  \\
V 
 & 1.05  & 1 & 0.992 &  & $+$ & $\infty$ &  & \multicolumn{1}{c}{$+\infty$} \\
VI  & 1.2  & 1 & \multicolumn{1}{c}{\tnote{*1}} &  & $0$ &  $(281)$ &  &  41.7 \\
VII  & 1.05  & 1 & 0.991 &  & $-$ & $117$ &  & \multicolumn{1}{c}{$-\infty$} \\
VIII  & 1  & 1 & 1 & $0$ & $0$ & $54$ & 55.2 & \multicolumn{1}{c}{$-\infty$}  \\
IX  & \multicolumn{1}{c}{$1/\beta$} & 0.992 & \multicolumn{1}{c}{\tnote{*2}} & $0$ & $0$ & $53$ & 49.1 & -1843.2 \\
\hline
\end{tabularx}
\begin{tablenotes}
\footnotesize
\item[Note:] For all cases 
I--IX, $k_0=150$.
$B=A(1- \theta)^{1- \theta} \theta ^{\theta}$ is the cost-function based productivity.
\item[*1]$\theta \approx 0.955392$ where ${1.2} (1- \theta)^{1- \theta} \theta ^{\theta}=B =1$.
\item[*2]$\theta \approx 0.998982$ where $(1/0.992) (1- \theta)^{1- \theta} \theta ^{\theta}=B =1$.
\end{tablenotes}
\end{threeparttable}
\end{center}
\end{table}

\section{Analysis}
\subsection{AK setting}
The AK model, which was formally developed by \citet{Frankel}, is one of the simplest fundamental models of endogenous growth.
Here, we employ this production model to study the population value function that discounts future generations to determine whether $N \to \infty$ is an optimal policy.
The optimal consumption path (\ref{optct}) for the AK setting with future discounting ($\theta=1$, $\beta<1$) may be specified as follows: 
\begin{align}
c_t^*[N]
= \left( \frac{{A} k_0}{S_{{N}-t}} \right)\left( \frac{S_{{N}-1}}{S_{N}} {A} \beta \right) \left( \frac{S_{{N}-2}}{S_{{N}-1}} {A} \beta \right) \cdots \left( \frac{S_{{N}-t}}{S_{{N}-t+1}} {A} \beta \right)
= \frac{({A} \beta)^t {A} k_0}{S_N}
\label{optlogct_ak}
\end{align}
The population value function (\ref{optval}), therefore, becomes:
\begin{align}
\mathcal{V}[N]
&=
\sum_{t=0}^{N} \beta^t \log 
\left( 
\frac{({A} \beta)^t {A} k_0}{S_N = 1 + \beta + \cdots + \beta^{N}} 
\right)
\notag \\
&=
\frac{\beta - ((1-\beta){N} + 1)\beta^{{N}+1}}{(1-\beta)^2} 
\log ({A} \beta)
+
\frac{1-\beta^{{N}+1}}{1-\beta} 
\log \left( \frac{1-\beta}{1-\beta^{{N}+1}}
{A} k_0
\right)
\label{optvctn_ak}
\end{align}
For the sake of the analysis, let us treat the planning horizon $N$ as a continuous variable and take the derivative with respect to $N$.
\begin{align}
\frac{ \mathrm{d} \mathcal{V}[N] }{\mathrm{d} N}
=
\lambda \left(
 \gamma - {N} \log ({A} \beta) + \log \left( 1-\beta^{N+1} \right)
\right)
\label{FOC}
\end{align}
where $\lambda$ and $\gamma$ are given as follows:
\begin{align*}
\lambda = \frac{\beta^{N} \log \beta^\beta}{1-\beta}>0,
&&
\gamma=1-\frac{\log ({A} \beta)^{1-\beta + \log \beta}}{\log \beta^{1-\beta}}-\log ({A} (1-\beta)k_0)
\end{align*}
To study the derivative sign of (\ref{FOC}), we consider the following two functions, whose equal values, i.e., $f[N]=g[N]$, convey the first-order condition of optimality.
\begin{align}
f[N]=-\gamma + N \log({A} \beta),  && g[N]=\log\left(1-\beta^{N+1}\right)
\label{fngn}
\end{align}

Before we proceed, let us examine the possible range of the parameter $A$.
The evaluation of the marginal product of capital (MPK) of our production function (\ref{PF}) net of depreciation, which should coincide with the real interest rate $\rho>0$, viz.,
\begin{align}
\frac{\partial Y_t}{\partial K_t} - \delta = {A}\theta \left(\frac{K_t}{L_t} \right)^{\theta-1} - \delta
= \rho >0
\label{mpk}
\end{align}
In AK models with complete capital depreciation, $\theta =\delta=1$ leads to ${A}=1 + \rho$.
Therefore, assuming that ${A}>1$ is relevant in this setting.\footnote{
Note also that $A\beta = (\delta+\rho)\beta$ indicates the discrepancy between the gross interest rate and generational preference rate of the population.
}

\begin{figure}[t!]
\centering
\includegraphics[width=0.495\textwidth]{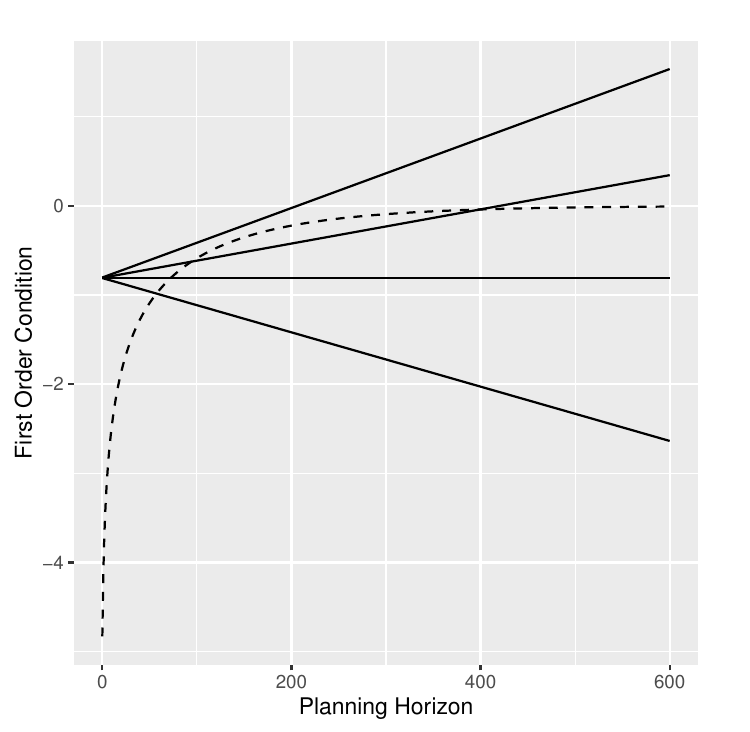}
\includegraphics[width=0.495\textwidth]{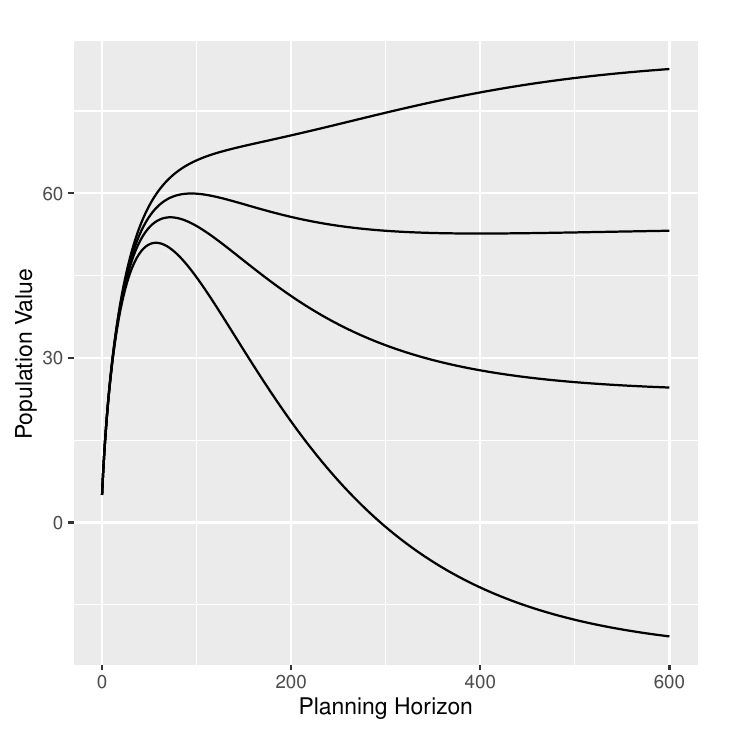}
\caption{
Left: The solid lines represent $g[N]$, as defined in (\ref{fngn}), for cases I--IV with clear correspondences, i.e., the steepest slope corresponds to case I, while case IV corresponds to the scenario with a negative slope.
The dashed line represents $f[N]$, which is also defined in (\ref{fngn}).
Right: The plots show population value functions $\mathcal{V}[N]$ for cases I--IV with obvious correspondences.
} \label{fandg}
\end{figure}

Figure \ref{fandg} (left) depicts the functions $f[N]$ and $g[N]$ under different parameters for cases I--IV.
Clearly, $f[N]$ is a linear function whose slope is $\log({A}\beta)$, and $g[N]$ monotonously increases and approaches zero, i.e., $g[\infty]=0$.
Here, we fix the discount factor $\beta$ and differentiate the productivity ${A}$ at four different levels.
As long as ${A} \leq 1/\beta$, so that $\log({A} \beta) \leq 0$, i.e., the slope is zero or negative, $f[N]$ will intersect with $g[N]$ at a single point, and the population value function $\mathcal{V}[N]$ has a single peak.
If ${A} > 1/\beta$ so that $f[N]$ has a positive slope, then $f[N]$ and $g[N]$ could either intersect with two points where the population value $\mathcal{V}[N]$ may rise, fall and then rise again, or never intersect so that the population value may rise indefinitely with respect to the planning horizon $N$.
Figure \ref{fandg} (right) depicts the population value functions for cases I--IV.

To visualize the optimal trajectory of the variables in various situations, we specify them here in the form of functionals.
By referencing (\ref{optlogct_ak}), the optimal trajectory of the undiscounted contribution to the population value becomes linear with respect to $t$, as follows:
\begin{align}
\log c_t^*[N] = \log \left( \frac{1-\beta}{1-\beta^{N+1}}{A} k_0 \right) + t \log ({A} \beta)
\label{optlogctn_ak}
\end{align}
In reference to (\ref{ctkt}) Appendix \ref{apd2}, the optimal trajectory of capital intensity becomes as follows:
\begin{align}
k_t^*[N] = \frac{S_{N-t}}{{A}}c_t^*[N]
= \frac{1-\beta^{N+1-t}}{1-\beta^{N+1}} ({A} \beta)^t k_0
\label{optktn_ak}
\end{align}

\begin{figure}[t!]
\includegraphics[width=0.329\textwidth]{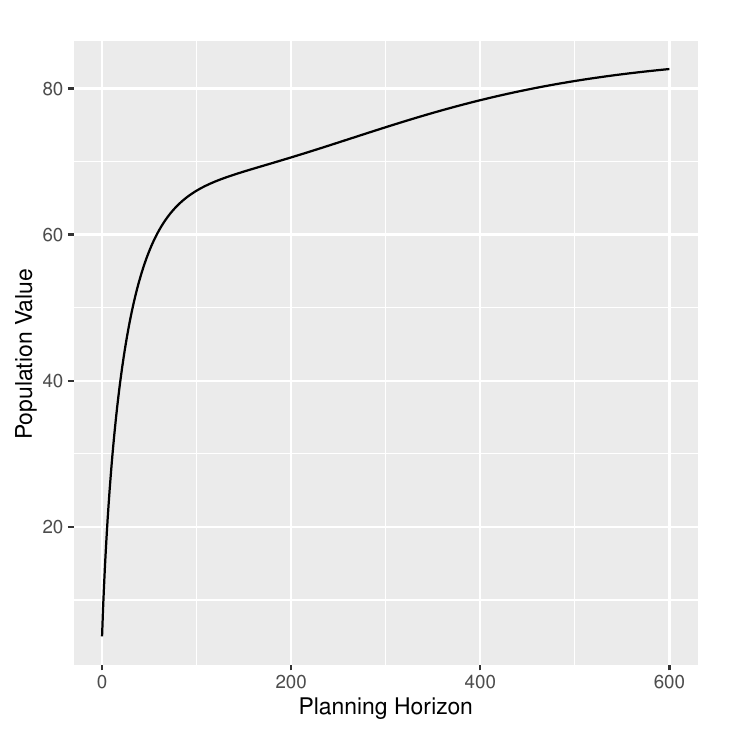}
\includegraphics[width=0.329\textwidth]{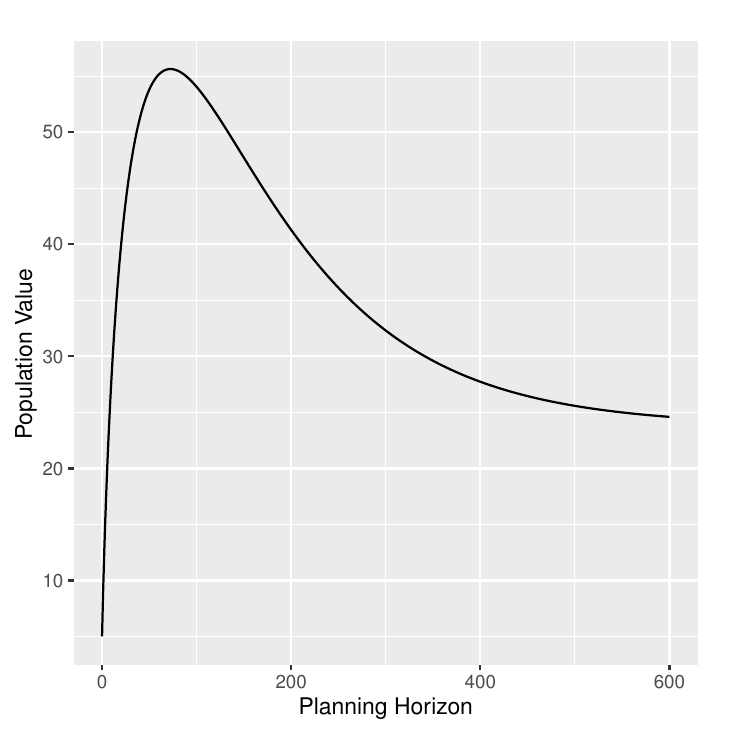}
\includegraphics[width=0.329\textwidth]{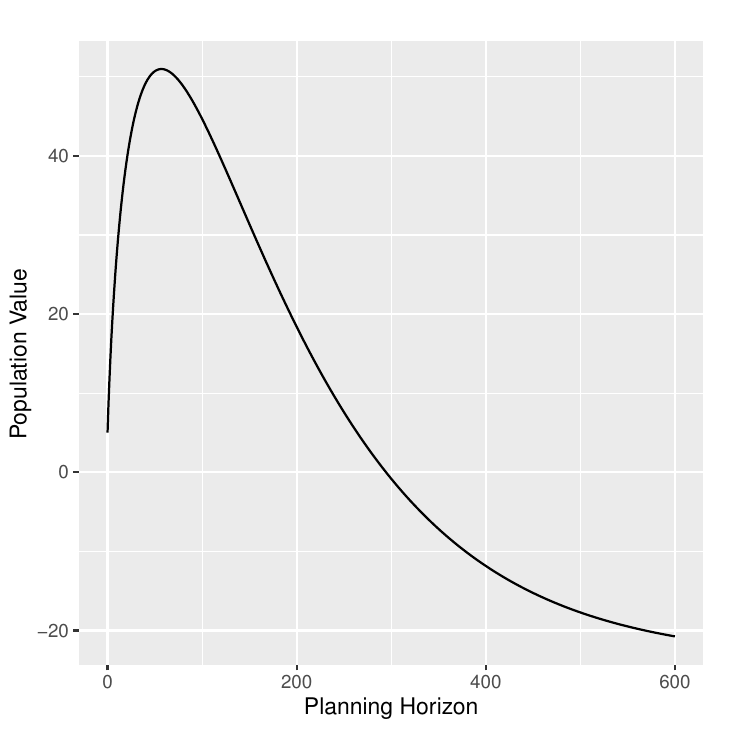}
\includegraphics[width=0.329\textwidth]{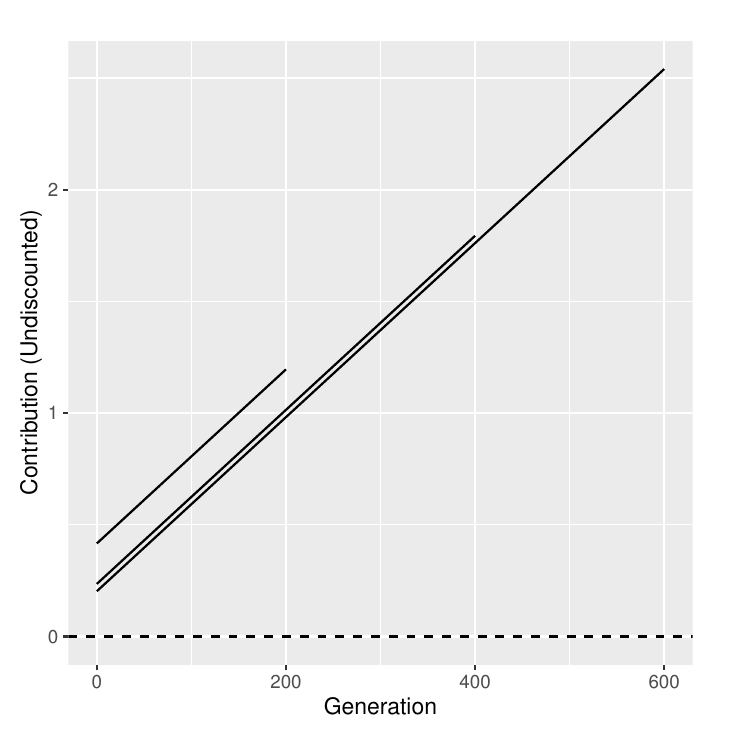}
\includegraphics[width=0.329\textwidth]{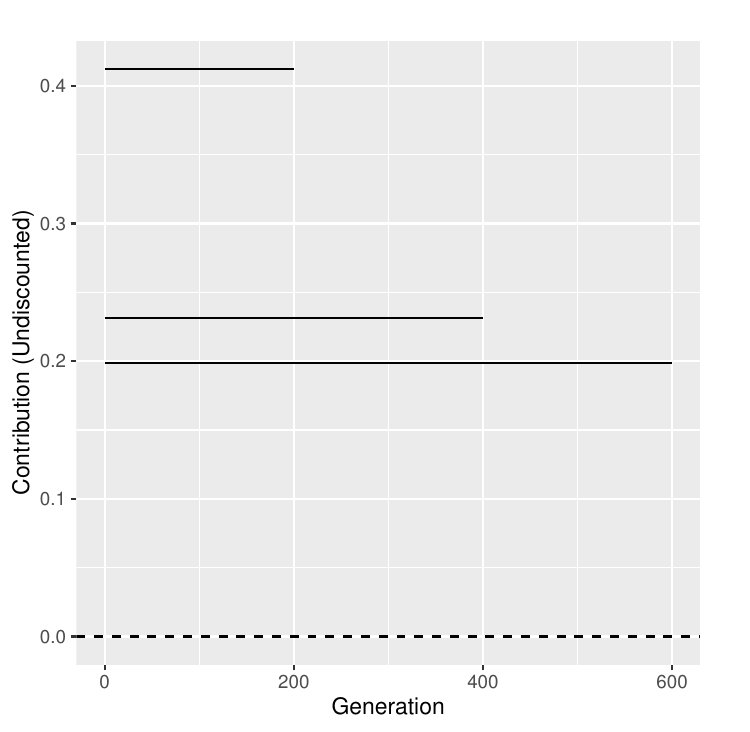}
\includegraphics[width=0.329\textwidth]{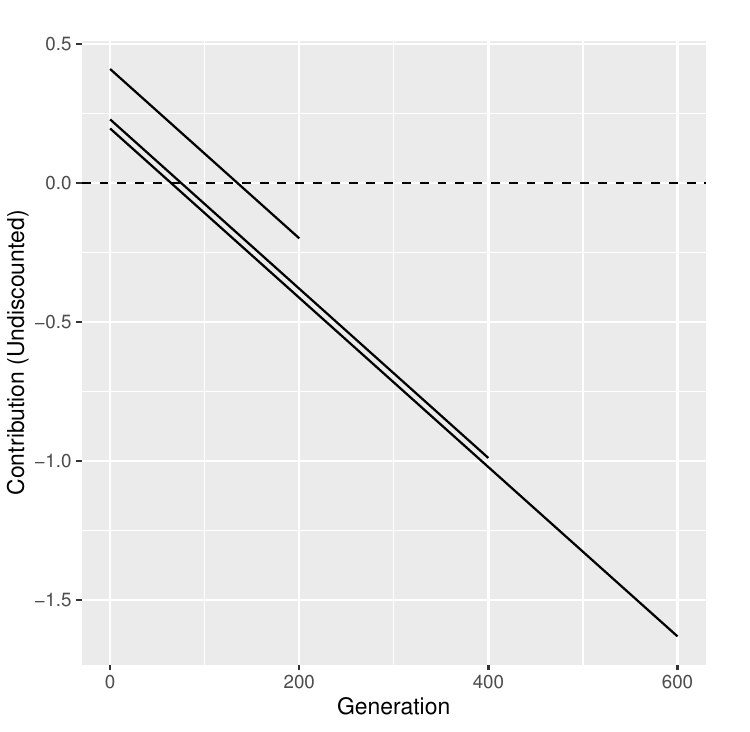}
\includegraphics[width=0.329\textwidth]{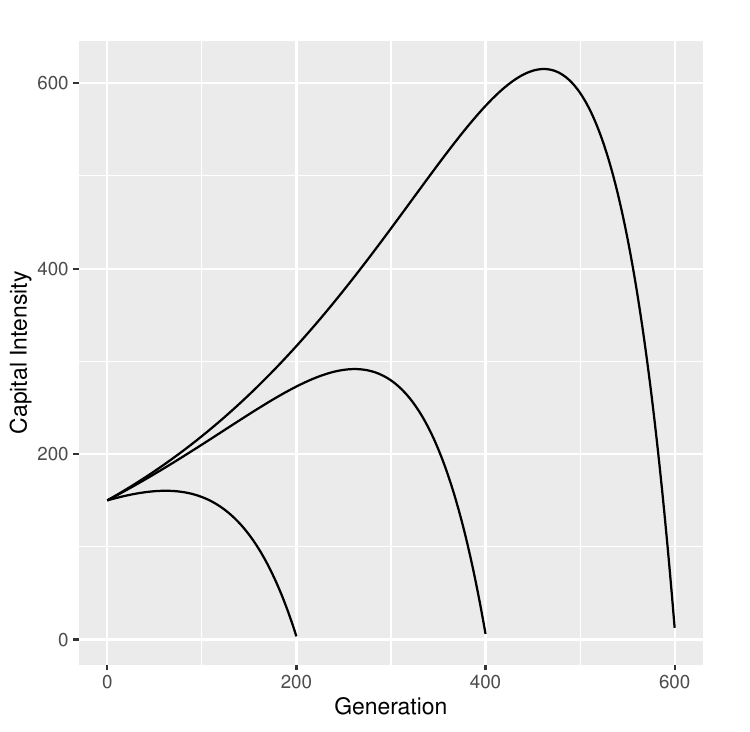}
\includegraphics[width=0.329\textwidth]{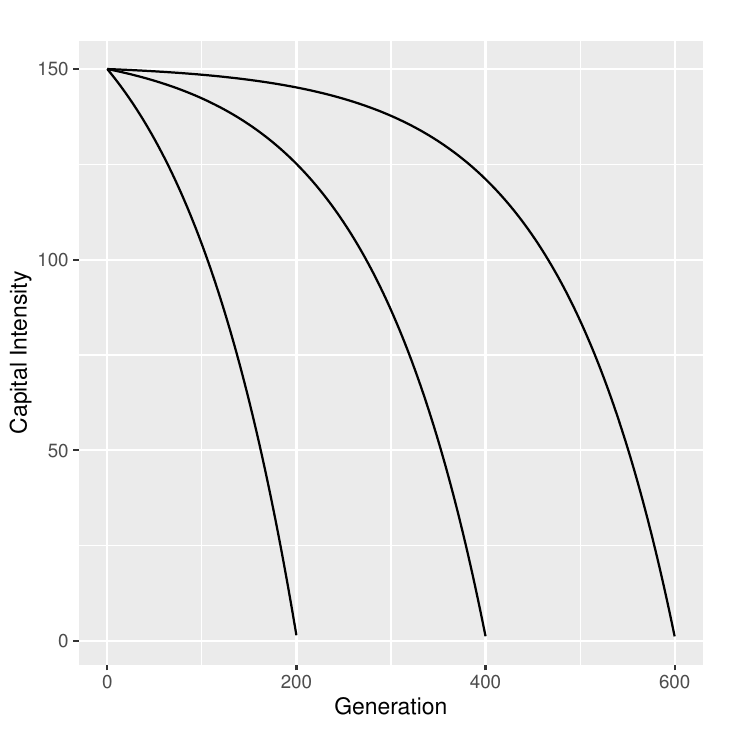}
\includegraphics[width=0.329\textwidth]{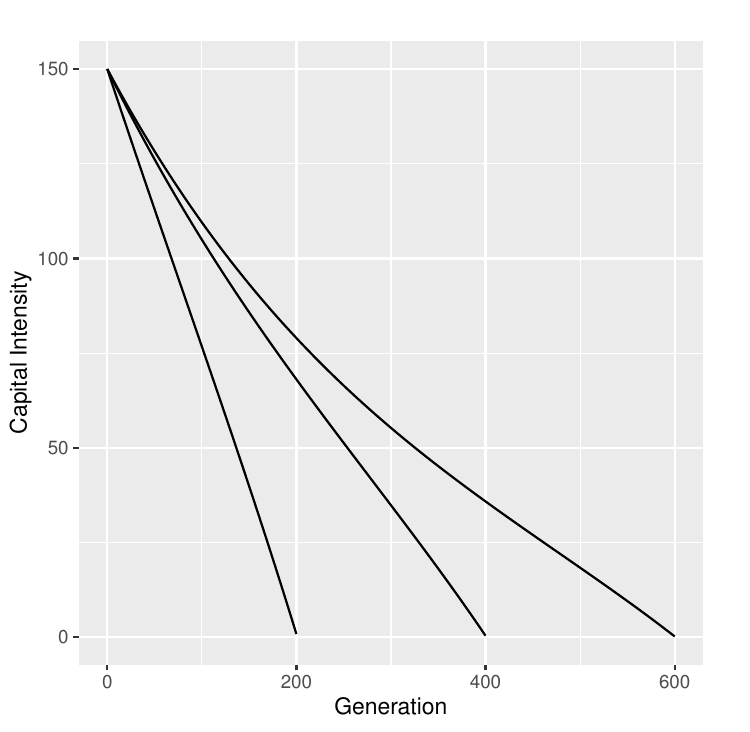}
\caption{
Population value function (top), seleted trajectories of undiscounted contribution (middle) and capital intensity (bottom), for AK setting cases I (left), III (center), and IV (right).
The trajectories are selected for $N=200, 400, 600$.
} \label{figxx}
\end{figure}

Figure \ref{figxx} displays the population value function $\mathcal{V}[N]$, specified as (\ref{optvctn_ak}), for cases I, III, and IV on the top.\footnote{These figures correspond to those depicted in Figure \ref{fandg} (right).
}
Note that, as long as $\beta<1$, the population value function will always converge to a finite value, regardless of the parameter settings in addition to $\beta$.
As we let $N\to\infty$ in (\ref{optvctn_ak}), we have the following:
\begin{align}
\mathcal{V}[\infty] = \frac{\log \left({A} (1-\beta)^{1-\beta} \beta^\beta \right)}{(1-\beta)^2} + \frac{\log (k_0)}{1-\beta}
\label{vctinfty_ak}
\end{align}
By comparing the population values obtained at the numerical solution of the first-order condition $f[N]=g[N]$, as referenced in (\ref{fngn}), with $\mathcal{V}[\infty]$ from (\ref{vctinfty_ak}), we know the optimal planning horizon $N^*$, which we display in Table \ref{tab_parameters} for cases I--IV.
The middle row of Figure \ref{figxx} displays the optimal trajectories of the undiscounted contributions to the population value, as described in (\ref{optlogctn_ak}), for planning horizons $N=200, 400, 600$ for cases I, III, and IV (from left to right).
Similarly, the bottom row of Figure \ref{figxx} displays the optimal trajectories of capital intensity, based on (\ref{optktn_ak}), for planning horizons $N=200, 400, 600$ for cases I, III, and IV (from left to right).

\subsection{ZD setting}
\begin{figure}[t!]
\centering
\includegraphics[width=0.329\textwidth]{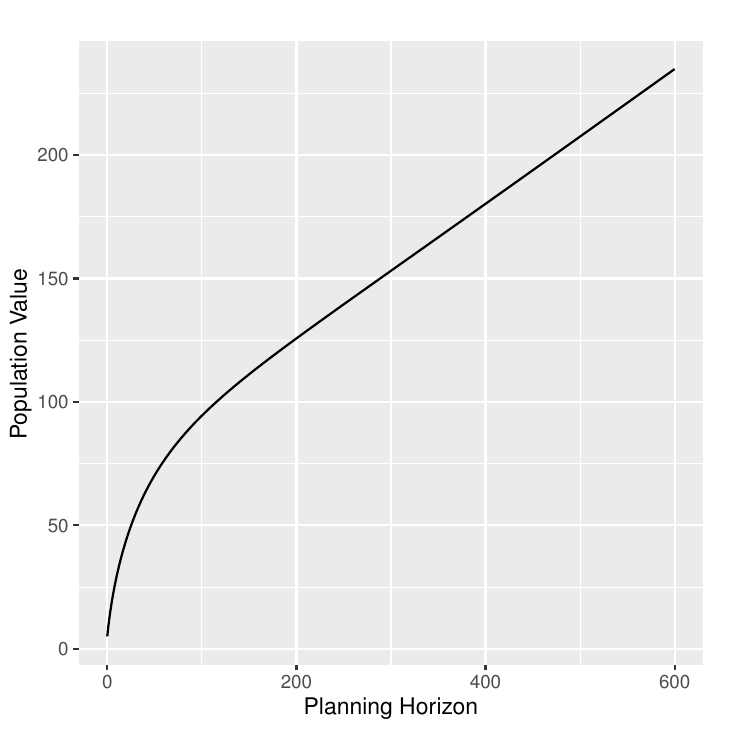}
\includegraphics[width=0.329\textwidth]{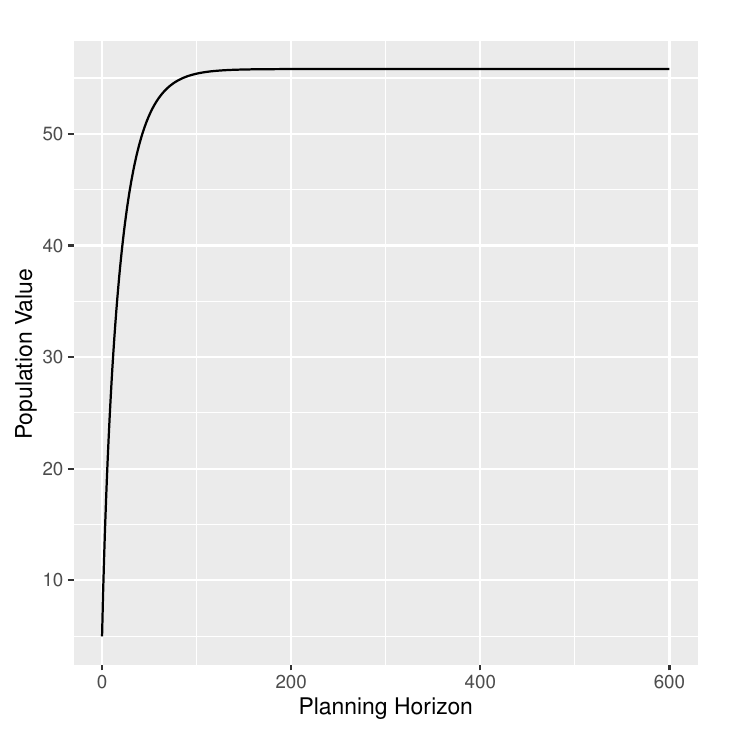}
\includegraphics[width=0.329\textwidth]{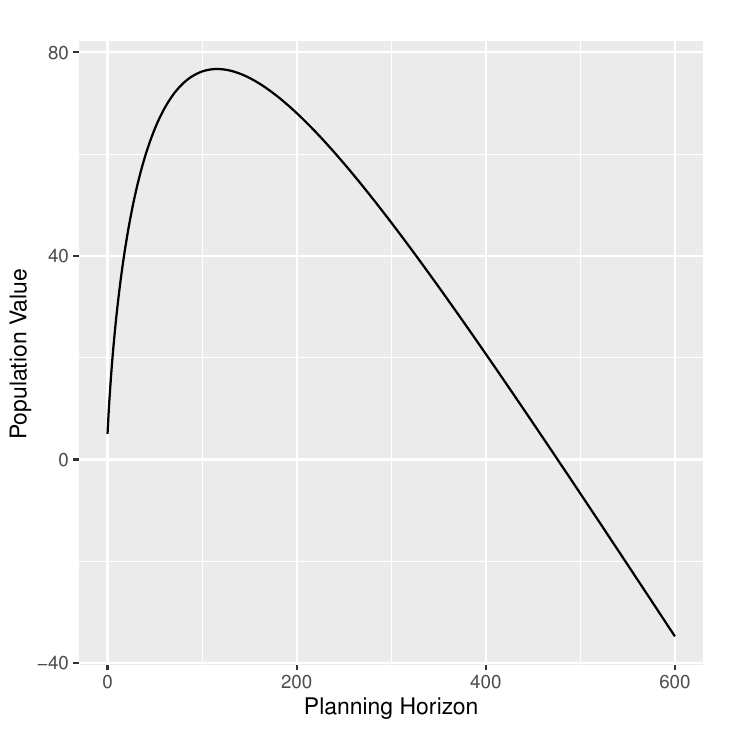}
\includegraphics[width=0.329\textwidth]{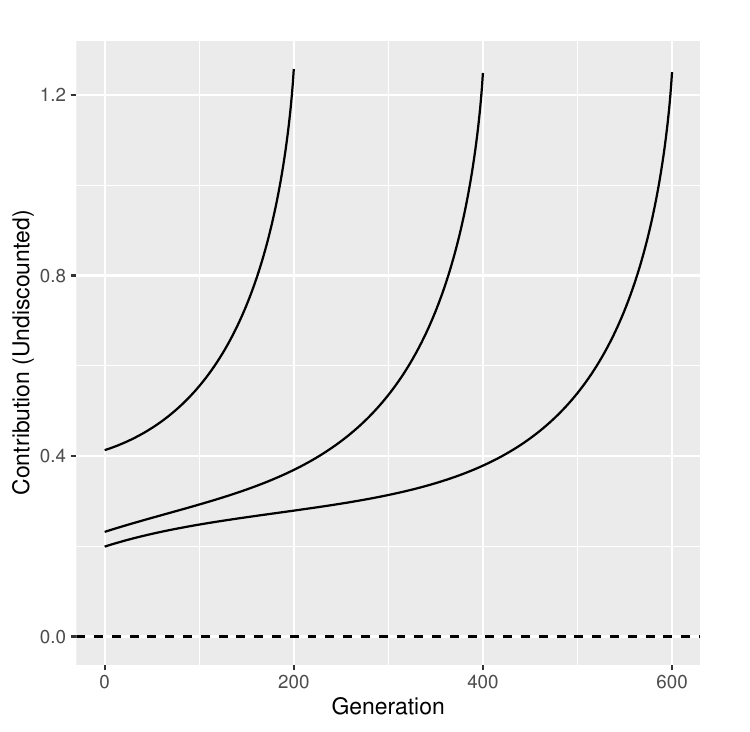}
\includegraphics[width=0.329\textwidth]{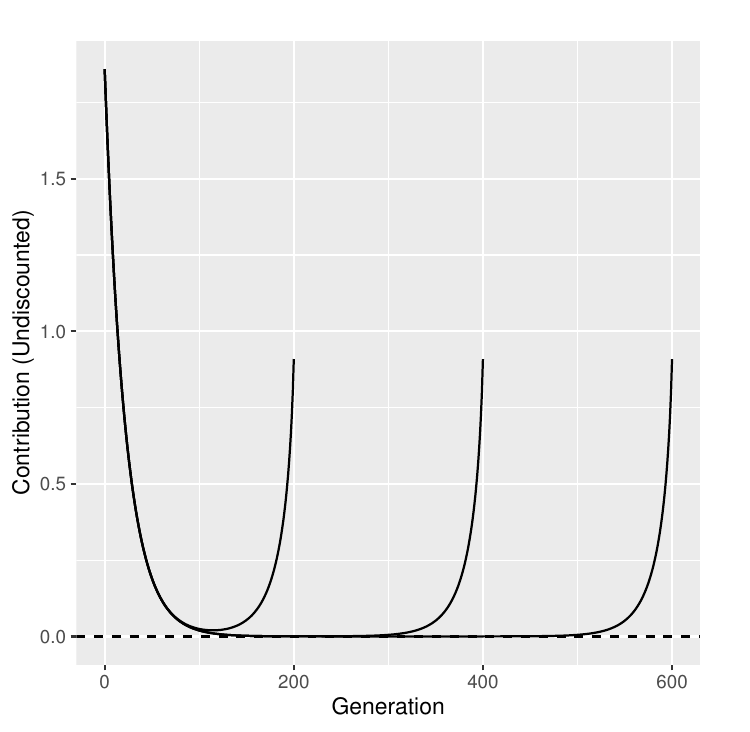}
\includegraphics[width=0.329\textwidth]{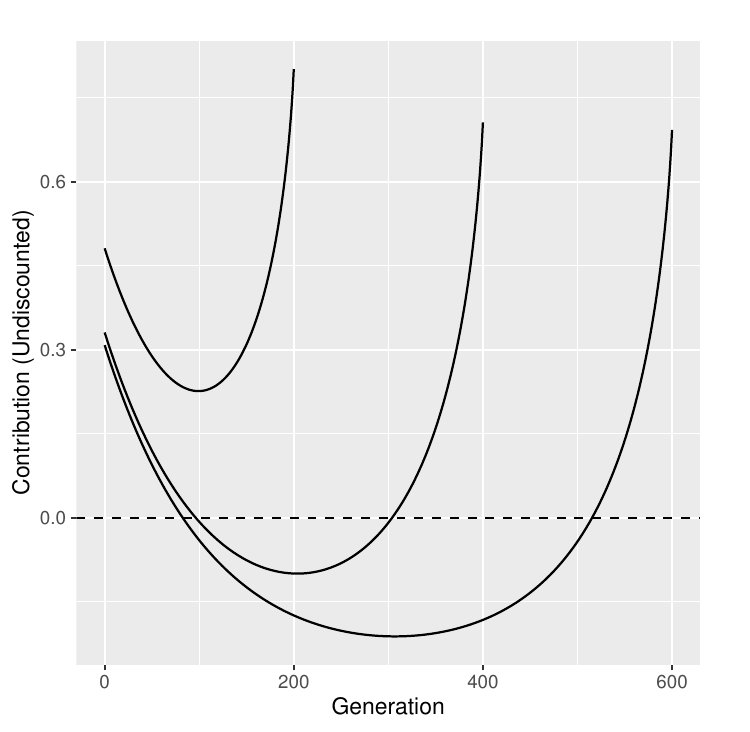}
\includegraphics[width=0.329\textwidth]{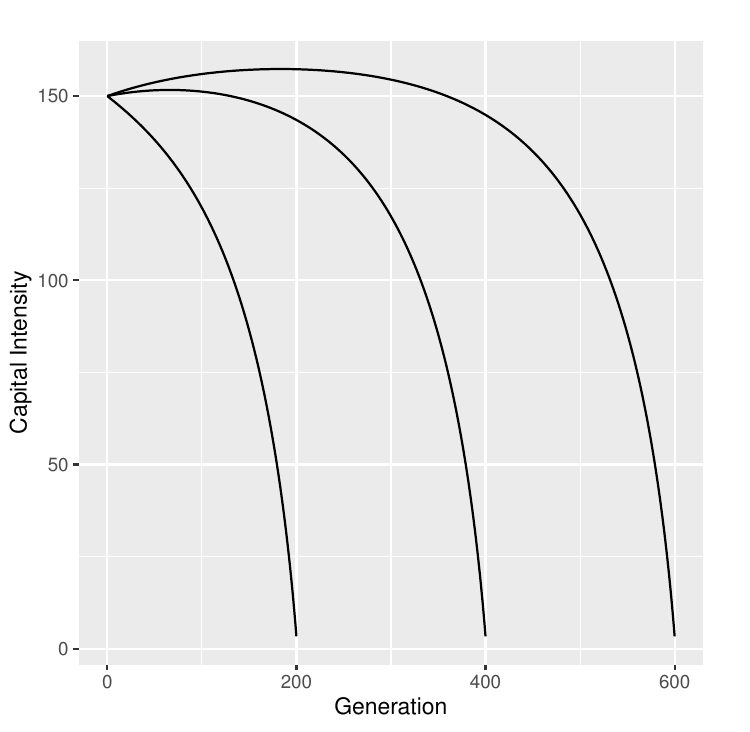}
\includegraphics[width=0.329\textwidth]{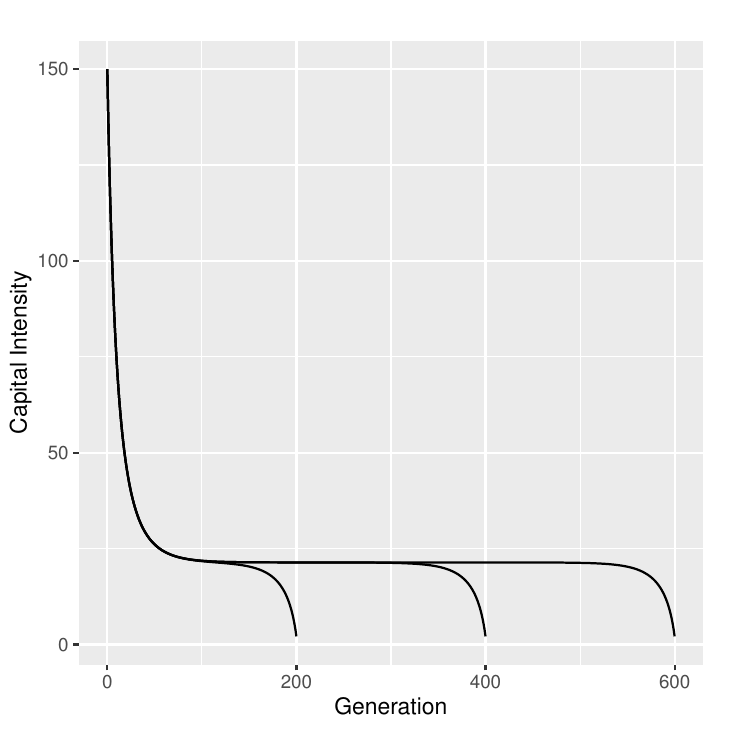}
\includegraphics[width=0.329\textwidth]{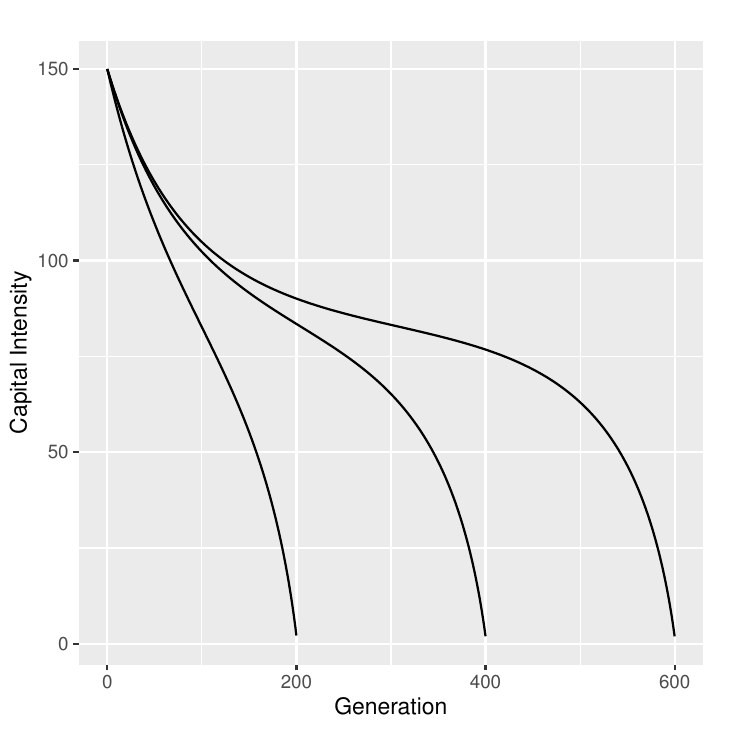}
\caption{
Population value function (top), selected trajectories of undiscounted contribution (middle) and capital intensity (bottom), for ZD setting cases V (left), VI (center), and VII (right).
The trajectories are selected for $N=200, 400, 600$.
Note that the population value function of the top-left panel is increasing indefinitely.
The parameters corresponding to each of the cases are given in Table \ref{tab_parameters}.
} \label{aaa}
\end{figure}

Here, we study ZD, i.e., $\beta =1$, under Cobb--Douglas production with an ordinary output elasticity of capital $\theta<1$.
The condition that $\beta\theta<1$ nonetheless allows us to evaluate the key summation as follows:
\begin{align*}
S_{\infty- \tau} = \lim_{N\to\infty} \frac{1-(\beta \theta)^{N-\tau+1}}{1-\beta\theta}
=\frac{1}{1-\beta\theta}
\end{align*}
We apply the above and $N\to\infty$ to (\ref{optct}) and obtain the following:
\begin{align}
\log c_t^*[\infty]
&=\log \left( \frac{{A} }{S_{{\infty}-t}} \right) + \theta \left( \sum_{i=1}^{t} \theta^{t-i} \log \left( \frac{S_{{\infty}-i}}{S_{{\infty}-i+1}} {A} \beta \theta \right) + \theta^{t}\log k_0 \right) \notag
\\
&=\log \left( {{A} }{(1-\beta\theta)} \right) + \left(\theta^{t-1} + \cdots +\theta + 1\right)\log({A}\beta\theta)^\theta + \theta^{t+1}\log k_0 \notag
\\
&= \frac{\log \left( {A} (1-\beta\theta)^{1-\theta} (\beta\theta)^{\theta} \right)}{1-\theta} 
+ \theta^{t} \left( \log (k_0)^{\theta} -\log ({A}\beta\theta)^{\frac{\theta}{1-\theta}} \right)
\label{optlnctinfty}
\end{align}
where the optimal trajectory for undiscounted contributions is geometrically convergent.
We can then apply $\beta=1$ to arrive at the following:
\begin{align*}
\log c_t^*[\infty] = \frac{\log \left( {A} 
(1-\theta)^{1-\theta} \theta^\theta 
\right)}{1-\theta} 
+ \theta^{t} \left( \log (k_0)^{\theta} -\log ({A}\theta)^{\frac{\theta}{1-\theta}} \right)
\end{align*}
The corresponding population value can hence be evaluated by the infinite sum of the undiscounted contributions, i.e.,
\begin{align*}
\mathcal{V}[\infty]
= \sum_{t=0}^\infty \log c_t^*[\infty] 
=& \frac{\infty \log \left( {A} 
(1-\theta)^{1-\theta} \theta^\theta 
\right) + \log (k_0)^{\theta} -\log ({A}\theta)^{\frac{\theta}{1-\theta}}}{1-\theta} \\
=&
\begin{cases}
+\infty & \iff  {A} (1-\theta)^{1-\theta} \theta^\theta = B >1 
\\
{\frac{\theta}{1-\theta}} \log \left(\frac{1-\theta}{\theta}k_0 \right) 
& \iff {A} (1-\theta)^{1-\theta} \theta^\theta = B = 1 
\\
-\infty  & \iff  {A} (1-\theta)^{1-\theta} \theta^\theta =B 
<1 
\end{cases}
\end{align*}
The above result indicates that if $B >1$, then the population value is ever increasing, and $N\to\infty$ must be the optimal solution.
However, if $B<1$, then $N\to\infty$ must not be optimal, and the population value must be maximized at a finite horizon $N^* \ll \infty$.
The case where $B =1$ is the knife-edge case in which the planning horizon $N$ does not matter (beyond a certain length) in maximizing the population value.
Case V corresponds to $B>1$, and $N^*=\infty$; case VII corresponds to $B<1$, and $N^* \ll \infty$;
and case VI corresponds to $B=1$, and $N^*$ is any number greater than $281$.\footnote{
The number is subject to decimal rounding.
}
Figure \ref{aaa} depicts the population value function $\mathcal{V}[N]$, along with the optimum trajectory of undiscounted contribution $\log c_t^*[N]$ and the capital intensity $k_t^*[N]$, given a planning horizon $N$, for sample ZD settings V, VI, and VII.
The final case VIII relates to the ZD/AK setting ($\theta=\beta=1$), where $B = A=1+\rho >1$, under complete depreciation, as described in (\ref{mpk}).
Thus, if $\rho>0$, then $A>1$, in which case the population value would increase without bound, as described above, leading to $N^*\to\infty$.
If, in turn, $A=1+\rho=1$, as in case VIII, the infinite horizon population value is evaluated as follows:
\begin{align*}
\mathcal{V}[\infty] = \lim_{\theta \to 1} {\frac{\theta}{1-\theta}} \log \left(\frac{1-\theta}{\theta}k_0 \right)  = -\infty
\end{align*}
This indicates that there exists a finite optimum horizon $N^* \ll \infty$.
Note that the population value function in this setting can be specified by applying a unitary discounting factor $(\beta\to 1)$ to the population value function (\ref{optvctn_ak}) of AK models as follows:
\begin{align*}
\mathcal{V}[N] &= \lim_{\beta\to 1} \left(
\frac{\beta - ((1-\beta){N} + 1)\beta^{{N}+1}}{(1-\beta)^2} 
\log ({A} \beta)
+
\frac{1-\beta^{{N}+1}}{1-\beta} 
\log \left( \frac{1-\beta}{1-\beta^{{N}+1}}
{A} k_0
\right)
\right)
\\
&= \lim_{\beta\to 1} 
\frac{1-\beta^{{N}+1}}{1-\beta} 
\log \left( \frac{1-\beta}{1-\beta^{{N}+1}}
k_0 \right)
=\lim_{\beta\to 1} (N+1)\beta^N \log \left( \frac{k_0}{(N+1)\beta^N} \right)
\\
&= (N+1) \log \left( \frac{k_0}{N+1} \right)
\end{align*}
The proof for the second identity, given $A=1$, is appended to Appendix \ref{apd3}.
The third identity is subject to L'H\^{o}pital's rule.
By the first-order condition $\frac{\partial \mathcal{V}[N]}{\partial N}=0$, the optimal planning horizon is evaluated as $N^* = \exp(-1+\log k_0) -1 \approx 54.182$.

\subsection{Intergenerational inequality}
\begin{figure}[t!]
\centering
\includegraphics[width=0.329\textwidth]{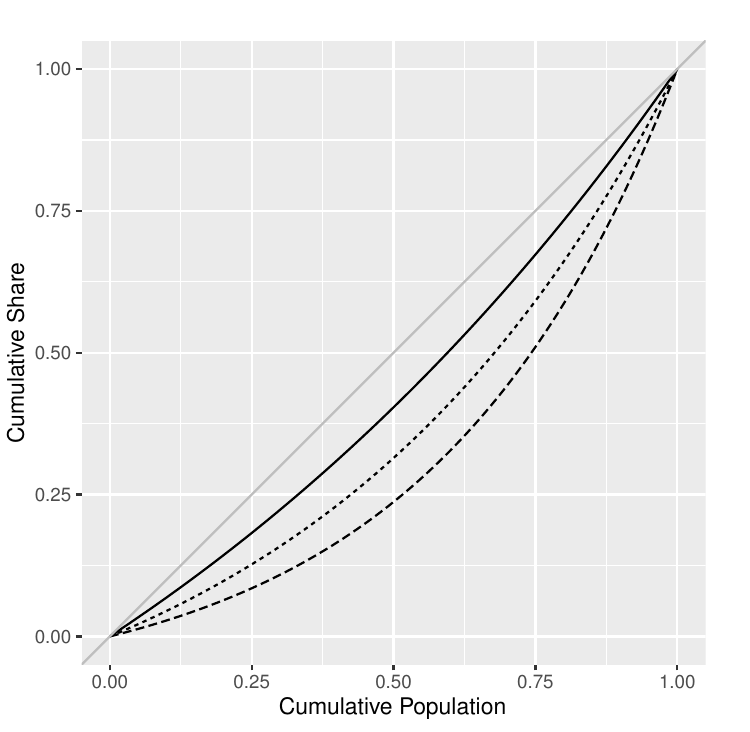}
\includegraphics[width=0.329\textwidth]{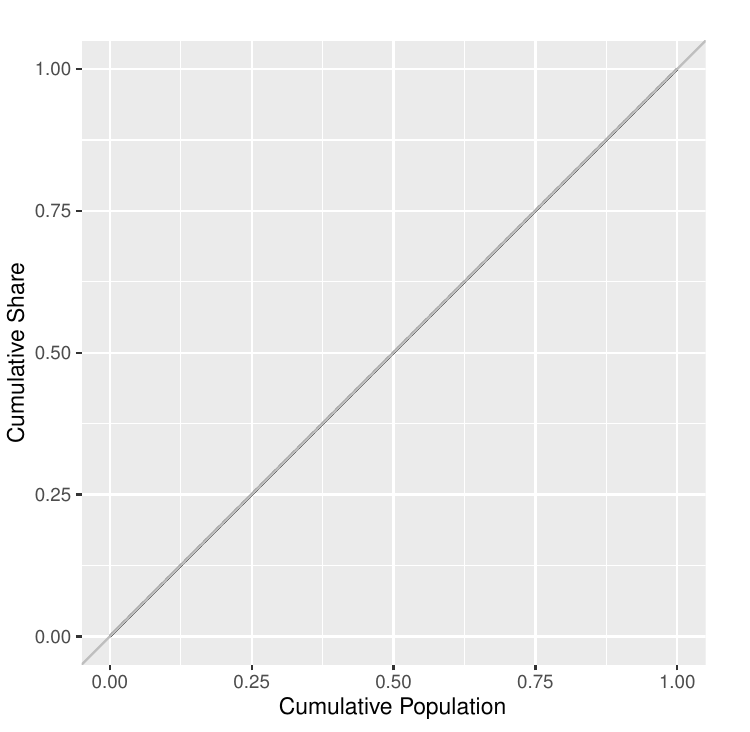}
\includegraphics[width=0.329\textwidth]{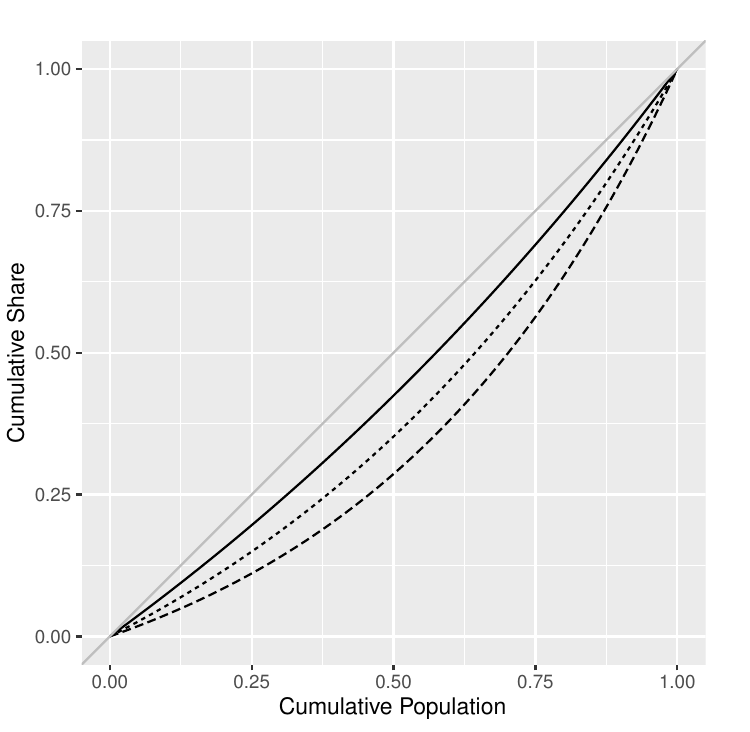}
\includegraphics[width=0.329\textwidth]{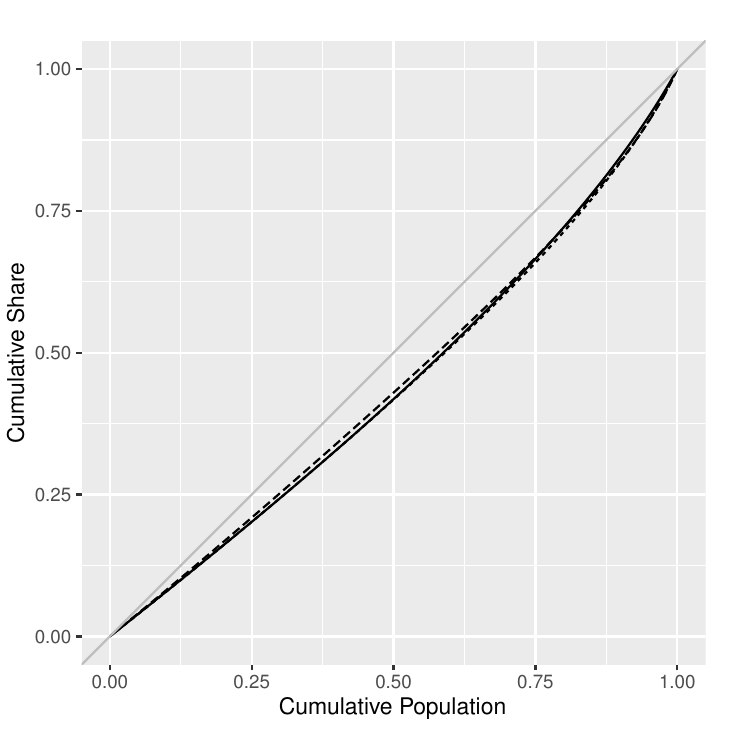}
\includegraphics[width=0.329\textwidth]{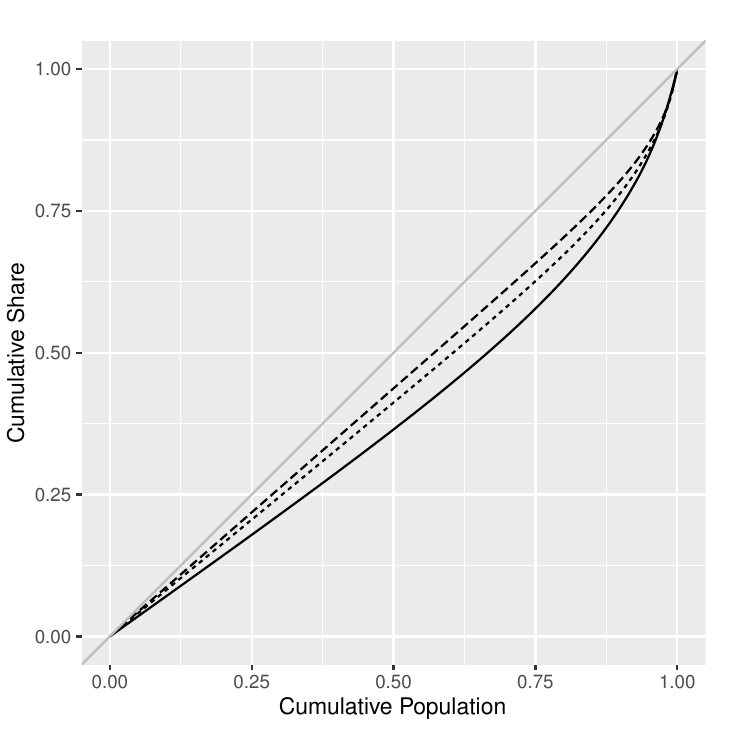}
\includegraphics[width=0.329\textwidth]{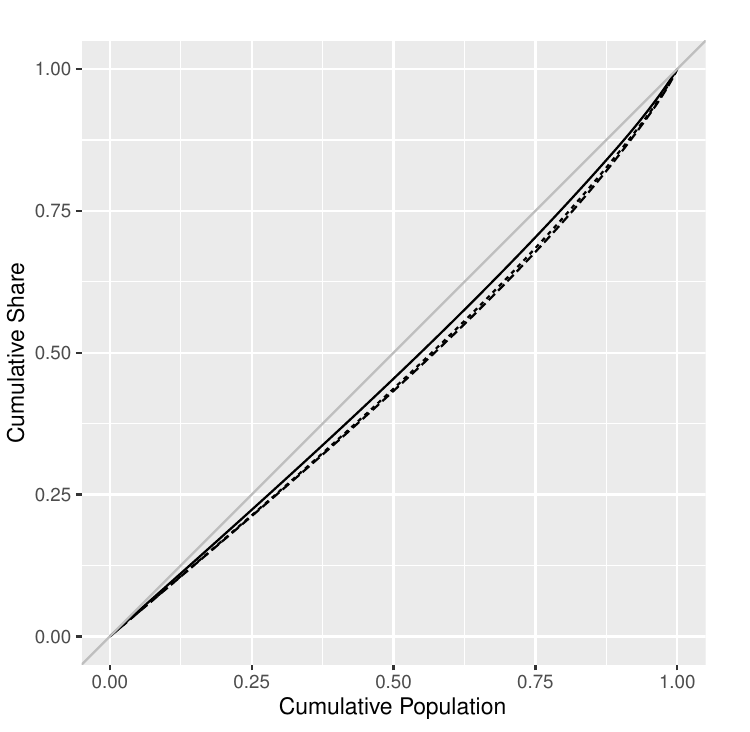}
\caption{Lorenz curves for $N=200$ (solid line), $N=400$ (dotted line), and $N=600$ (dashed line).
The top row panels (from left to right) correspond to AK settings I, III, and IV.
The second row panels (from left to right) correspond to ZD settings V, VI, and VII.
} \label{aaae}
\end{figure}

Figure \ref{figxx} (middle row) shows that consumption inequality across generations increases as the planning horizon extends under future discounting (except in the knife-edge case III when $\log (A\beta)=0$), whereas it seems to decrease in Figure \ref{aaa} (middle row) for cases with ZD.
To validate this conjecture, Figure \ref{aaae} shows how the Lorenz curve shifts with respect to the planning horizon (namely, $N=200, 400, 600$) for AK setting cases I, III, and IV and ZD setting cases V, VI, and VII.
These Lorenz curves (of Figure \ref{aaae}) are based on the sequence of optimal consumption levels for a given planning horizon, i.e., $c_t^*[N]$.
The inequality level increases in AK settings as the planning horizon expands.
On the other hand, inequality level is relatively insensitive to the planning horizon for ZD setting cases, except for case VI, when $\log B =0$, where inequality decreases as the planning horizon increases.
\begin{figure}[t!]
\centering
\includegraphics[width=0.49\textwidth]{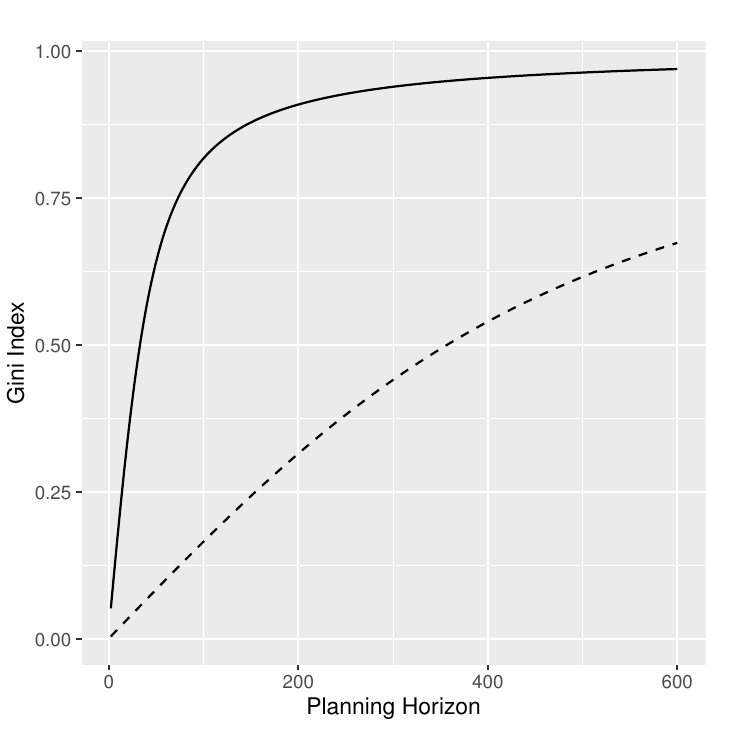}
\includegraphics[width=0.49\textwidth]{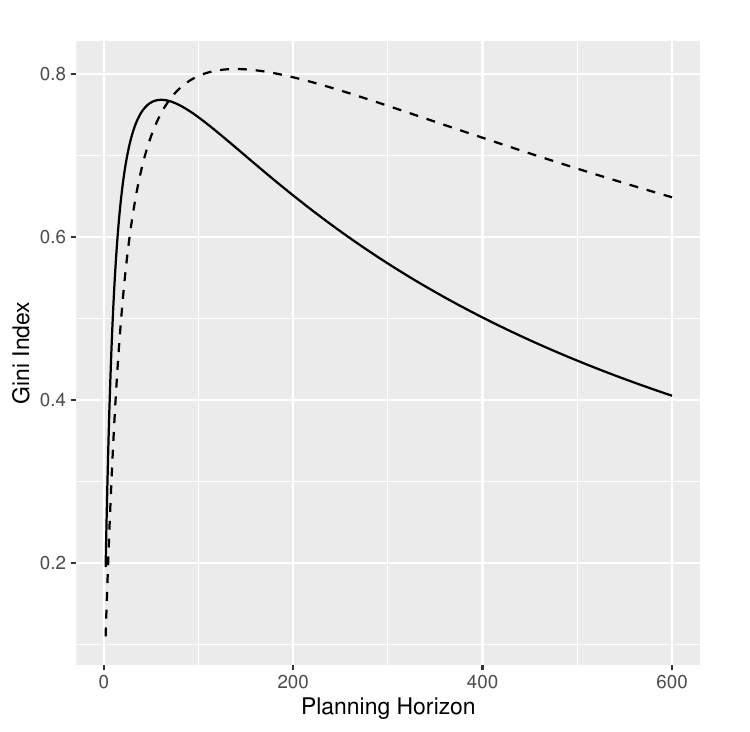}
\caption{
The panels depict the planning horizon vs Gini Index, i.e., $(N, \mathcal{G}[N])$ for AK setting cases with different discount rates (left), and ZD setting cases with different output elasticities of capital (right).
The parameter settings for the left panel are $\theta=1$, $k_0=150$, $A=1$, $\beta=0.9$ (solid line), and $\beta=0.99$ (dashed line).
The parameter settings for the right panel are $\beta=1$, $k_0=150$, $A=1$, $\theta=0.9$ (solid line), and $\theta=0.95$ (dashed line).
} \label{gini}
\end{figure}

For further analysis, let us consider the Gini index, a popular measure of inequality, defined on the basis of the optimal stream of consumption $c_t^*[N]$, 
as follows:
\begin{align*}
\mathcal{G}[N] = \frac{ \sum_{t^\prime=0}^N  \sum_{t=0}^N \left|
c_{t}^*[N] - c_{t^\prime}^*[N] \right| }{2N \sum_{t=0}^N c_t^*[N]}
\end{align*}
Figure \ref{gini} shows the Gini index $\mathcal{G}$ as a function of the planning horizon $N$ under various parameter settings.
For both panels, the underlying parameters are fixed at $k_0=150$ and $A=1$.
The left panel corresponds to the AK settings ($\theta=1$) as the discount rate is increased from $\beta=0.9$ (solid line) to $\beta=0.99$ (dashed line).
In the AK settings, consumption tends to become more unequal across generations as the planning horizon expands.
However, this effect is mitigated when future discounting decreases.
In contrast, the right panel indicates that ZD settings tend to equalize consumption across generations as the planning horizon expands, whereas inequality is enhanced by the higher output elasticity of capital.

It is important to note that the distributional analysis and the resulting intergenerational inequality trends presented here rely specifically on the logarithmic utility specification ($\gamma = 1$). The dynamics of intergenerational inequality might differ under general CRRA preferences, and our normative statements regarding inequality should be interpreted within the context of this specific functional form.

\subsection{Initial action}
Let us focus on the initial action of the optimal consumption schedule, which we specify as follows based on (\ref{optct}) or (\ref{ctkt}): 
\begin{align}
c_0^*[N] = \frac{{A}(k_0)^\theta}{S_{N}} = \frac{1-\beta\theta}{1- (\beta\theta)^{N}} {A}(k_0)^\theta
\label{ini}
\end{align}
We take the derivative of (\ref{ini}) and obtain the following result:
\begin{align*}
\frac{\mathrm{d} c_0^*[N]}{\mathrm{d} N} = \frac{A(k_0)^\theta \left(1 - \beta\theta \right)(\beta\theta)^N \log(\beta\theta)}{\left(1 - (\beta\theta)^N \right)^2} <0
\end{align*}
since $\beta\theta<1$.
In other words, the more future generations are treated equally, the more current generation must reduce their consumption.
In any case, the initial action monotonically decreases in the planning horizon and ultimately converges to the following:
\begin{align}
c_0^*[\infty] = (1- \beta\theta){A}(k_0)^\theta
\label{iniinfty}
\end{align}
By taking its derivative with respect to $\beta$, we have:
\begin{align*}
\frac{\partial c_0^*[\infty]}{\partial \beta} = -\theta A (k_0)^\theta < 0
\end{align*}
In other words, the more we consider future generations (by raising $\beta$), the more the current generation must reduce their consumption.
From another perspective, we can solve equation $c_0^*[N] = \nu$ for $N$ with respect to (\ref{ini}), where $\nu$ is the well-being subsistence, as follows:
\begin{align*}
N = \frac{\log\left( (1-\beta\theta) A (k_0)^\theta \right) - \log \nu}{\log (\beta\theta)}
\end{align*}
This $N$ is the subsistence-proof size of potential generations.
Finally, for AK models, the optimal initial action by (\ref{iniinfty}) is $c_0^*[\infty] = (1-\beta)Ak_0$, but this value approaches zero if future generations are given equal consideration as the current generation ($\beta \to 1$).
That is, in AK models with an infinite planning horizon, giving ultimate consideration to future generations requires the current generation to ultimately reduce their consumption, i.e., $c_0^*[\infty] \to {\nu}$.
Alternatively, if we relax the infinite horizon assumption, the initial action for AK models with ultimate consideration for the future generations becomes:
\begin{align*}
c_0^*[N] = \lim_{\beta \to 1} \frac{1-\beta}{1- \beta^{N}} {A}k_0
=\frac{Ak_0}{N}
\end{align*}
and we are left with the AK/ZD version of subsistence-proof size of potential generations, that is, $N=Ak_0/\nu$.

\section{Concluding Remarks}

This paper has investigated the optimal longevity of a dynasty under a Critical-Level Utilitarian framework.
By treating the planning horizon as an endogenous choice variable within a dynamic programming setting, we derived closed-form analytical solutions that link economic fundamentals to the ethical duration of a society.
Our analysis highlights two distinct regimes determined by productivity and time preference.
In the AK setting with discounting, the sign of $\log(A\beta)$ serves as the conservative boundary determining whether the optimal horizon is finite or infinite.
Similarly, in the Zero Discounting (ZD) setting, which prioritizes intergenerational equity, the sign of the cost-based productivity $\log(B)$ dictates the optimal longevity.
These results demonstrate that under low productivity or strict ethical constraints, a finite horizon is not an anomaly but a structurally optimal response to avoid negative social welfare.

Our findings also reveal a striking contrast in intergenerational inequality between the two settings.
In the AK setting, a longer planning horizon is associated with greater inequality across generations, although a higher discount factor $\beta$ mitigates this effect.
Conversely, in the ZD setting, extending the horizon tends to reduce inequality, as the burden of capital accumulation is shared more evenly.
However, a higher output elasticity of capital $\theta$ in the ZD model exacerbates long-term inequality.
These observations suggest that the pursuit of longevity imposes different distributional costs depending on the underlying technology and ethical discount rates.
However, these specific normative observations regarding intergenerational inequality are closely tied to the logarithmic utility specification ($\gamma = 1$) and should be interpreted with this limitation in mind.

A key theoretical insight emerges from the behavior of the initial consumption $c_0$ as we approach the limit of perfect intergenerational impartiality ($\beta \to 1$).
In both models, a higher $\beta$ generally requires the current generation to reduce consumption to support a longer dynasty.
In the limiting case of the infinite-horizon AK model with $\beta \to 1$ (and $\theta=1$), the optimal initial consumption approaches zero (or the subsistence level).
This implies that if the future holds infinite potential, the current generation is ethically compelled to make the \emph{ultimate sacrifice} for the sake of an eternal dynasty.
In contrast, in a stagnant economy where such sacrifice yields diminishing returns, our model prescribes termination.
Thus, our framework illuminates an ethical asymmetry: it demands immense endurance when the future is promising, but offers the ``mercy'' of termination when the future promises only misery.

Methodologically, our approach ensures time consistency through the recursive structure of dynamic programming.\footnote{
Formally, suppose that at some intermediate period $t < N^*$, the planner re-optimizes and chooses a different horizon $N^\# \neq N^*$ (and a corresponding consumption path) that yields a higher continuation value than the original plan.
This would imply that the original total value could have been improved by switching to this new path at $t$, contradicting the optimality of the original plan $(c_0^*, \dots, c_{N^*}^*)$.
Therefore, the optimal horizon $N^*$ determined at $t=0$ remains optimal for any subsequent generation $t$.
}
Unlike ad hoc termination rules, the finite horizon derived here is the result of a consistent optimization process where the continuation value falls below the critical threshold.
We acknowledge, however, that our model relies on a logarithmic utility function and a fixed population size per generation to maintain analytical tractability.
Introducing more complex dynamics, such as non-linear production technologies under general CRRA preferences, endogenous population size, or environmental constraints, would likely preclude closed-form solutions and require numerical approaches.
Exploring these complex dynamics remains a promising avenue for future research.

Finally, our results offer a normative perspective on the demographic trends observed in modern mature economies.
The choice to limit the number of offspring---or in our macroeconomic interpretation, to limit the longevity of the dynasty---is often viewed as a failure of vitality.
However, from the perspective of Critical-Level Utilitarianism, this may be interpreted as an \emph{altruistic termination}.
If the current generation anticipates that economic or environmental constraints will condemn future generations to lives barely worth living, choosing a finite horizon is a rational and ethical decision to prevent the creation of suffering.

\appendix
\normalsize
\renewcommand{\thesection}{\Alph{section}}

\makeatletter
\renewcommand{\@seccntformat}[1]{Appendix \csname the#1\endcsname\quad}
\makeatother

\numberwithin{equation}{section}

\section{} \label{apd1}
\begin{figure}[t!]
\centering
\centering
\includegraphics[width=0.85\textwidth]{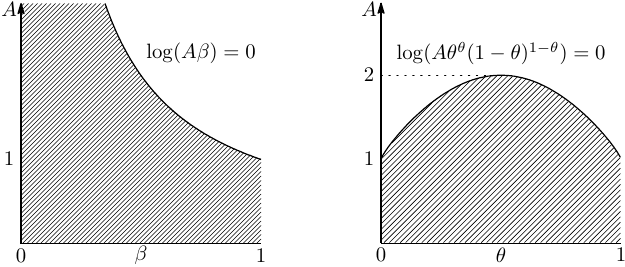}
\caption{
The shaded area (marginal lines not included) corresponds to the set of parameters $(\beta, A)$ where $A\beta <1$ (left), and $(\theta, A)$ where $B=A\theta^\theta(1-\theta)^{1-\theta}<1$ (right).
}  \label{Fig2}
\end{figure}

Below, we write a Cobb--Douglas production function and its dual unit cost function:
\begin{align*}
{Y} = A {K}^{\theta} {L}^{1- \theta},
&&
p = {B}^{-1} {r}^\theta w^{1-\theta} 
\end{align*}
where $p$, $r$, and $w$ denote prices corresponding to $Y$, $K$, and $L$, respectively.
The remaining parameter $A$ is referred to as the productivity, and $B$ is the cost function-based productivity.
Applying Shephard's lemma on the dual function leads to the following.
\begin{align*}
\frac{\partial p}{\partial r} = \frac{\theta}{B}\left( \frac{r}{w} \right)^{\theta -1} = \frac{K}{Y},
&&
\frac{\partial p}{\partial w} = \frac{1-\theta}{B}\left( \frac{r}{w} \right)^{\theta} =\frac{L}{Y}
\end{align*}
On the basis of these equations, the marginal product of capital (MPK) can be readily evaluated as follows:
\begin{align*}
\text{MPK} = \frac{\partial Y}{\partial K} = A\theta \left( \frac{K}{L} \right)^{\theta -1}
&= A \theta^\theta (1-\theta)^{1-\theta} \left( \frac{r}{w} \right)^{\theta-1}
\\
A\theta \left( \frac{K}{L} \right)^{\theta -1}
&=\frac{Y}{K}\theta
= B \left( \frac{r}{w} \right)^{\theta-1}
\end{align*}
where $r/w \in (0, \infty)$ is the marginal rate of substitution (MRS) between capital and labor.
By comparison, we are left with $B= A \theta^\theta (1-\theta)^{1-\theta} $.
Moreover,
\begin{align*}
A = \frac{
\text{MPK}}{\theta^\theta (1-\theta)^{1-\theta} } 
\left( \frac{r}{w} \right)^{\theta-1},
&&
B = \text{MPK} \left( \frac{r}{w} \right)^{\theta-1}
\end{align*}
On the other hand, we recall the breakdown equation of the total output (\ref{balance}) and take the partial derivative as follows:
\begin{align}
\frac{\partial Y_t}{\partial K_t} = \left( \frac{\partial K_{t+1}}{\partial K_{t}} - 1 \right) + \delta = \rho + \delta = \text{MPK}
\end{align}
where $\rho >0$ denotes the rate of interest.
Hence, if $\theta =1$ (AK setting), $A=B=\text{MPK}=\rho+\delta$, and if $\delta=1$ (complete depreciation), it must be the case that $A=B=1+\rho>1$.
From the study in Section 3.1, we note that the condition $A\beta <1$ implies the nonincreasing property of undiscounted contributions (or utilities less the critical level) with respect to generations.
The knife-edge case where $A\beta =1$ implies that the undiscounted contributions are constant over generations.
Figure \ref{Fig2} (left) depicts the set of possible values for $(\theta, A)$ that satisfies $A\beta < 1$.
Additionally, from the study in Section 3.2, we note that the condition $B<1$ implies the existence of a finite optimal planning horizon.
Otherwise, if $B>1$, the optimal planning horizon will be infinite.
The knife-edge case where $B=1$ implies that the optimal planning horizon is indeterminate.
Figure \ref{Fig2} (right) depicts the set of possible values for $(\theta, A)$ that satisfies $B < 1$.

\section{} \label{apd2}
\begin{lemma} \label{p1}
The value function of the Bellman equation (\ref{Bellman}) is as follows:
\begin{align*}
\mathcal{V}_{t}\left[ k_t;
N\right] & =  S_{{N}-t} \theta \log (k_t) + {R}_t
\end{align*}
where $S_{N-t}$ is specified by (\ref{sum}) and ${R}_t$ is a term that does not depend on $k_t$.
\end{lemma}
\begin{proof}
We show this by induction.
First, the value function at $t = {N}$ is evaluated via (\ref{Bellman}):
\begin{align*}
\mathcal{V}_{N} \left[k_{N};
{N} \right] = \max_{c_{N}} \left( \log c_{N} + \beta \mathcal{V}_{N+1}[k_{{N}+1}= {A} (k_{N})^\theta - c_{N}; {N}] \right)
\end{align*}
Since $k_{{N}+1}=0$ by (\ref{const}), $c_{N} = {A} (k_{N})^\theta$ must be true.
Additionally, $\mathcal{V}_{{N}+1}=0$ must be true for efficiency.
Thus, the final maximization is bounded, i.e.,
\begin{align*}
\mathcal{V}_{N}\left[ k_{N};
{N} \right] = \log c_{N} = \log {A} + \theta \log k_{N}
\end{align*}
Because $S_{N-N}=S_{0}=1$ and since ${A}$ is a constant, the lemma holds true for $t={N}$.
Suppose that the lemma holds true for $t+1$.
Then,
\begin{align*}
\mathcal{V}_{t}\left[k_t; N \right] 
= &\max_{c_{t}} \left( \log c_{t} + \beta \mathcal{V}_{t+1} \left[k_{t+1} = {A} (k_t)^\theta - c_t ; N \right] \right) \\
=&\max_{c_{t}} \left( \log c_{t} + \beta \left( S_{{N}-t-1} \theta \log \left( {A} (k_t)^\theta - c_t \right) + {R}_{t+1} \right) \right) 
\end{align*}
Below are the corresponding first-order condition and its solution:
\begin{align}
\frac{1}{c_{t}} - \frac{(\beta\theta)S_{{N}-t-1}}{ {A} (k_t)^\theta - c_t }=0,
&&
\text{or,}
&&
c_{t} = \frac{{A} (k_t)^\theta}{S_{{N}-t}}
\label{ctkt}
\end{align}
Here, we use $(\beta \theta)S_{{N}-t-1}=\beta\theta + (\beta\theta)^2 + \cdots + (\beta\theta)^{N-t}=S_{N-t}-1$.
By plugging the above solution back into the maximand, we arrive at the following result:
\begin{align*}
\mathcal{V}_t [k_t;
N] 
&= \log \left( \frac{{A} (k_t)^\theta}{S_{{N}-t}} \right)
+ \beta \left(
R_{t+1} + S_{N-t-1} \theta \log \left(
{A} (k_t)^\theta - \frac{{A} (k_t)^\theta}{S_{{N}-t}}
\right)
\right)
\\
&=
S_{{N}-t} \theta \log (k_t) +
\left(
\beta {R}_{t+1} +
S_{{N}-1} \log \left( \frac{{A}}{S_{{N}-t}} \right)
+  \log (S_{N-1}-1)^{S_{N-1}-1} 
\right) 
\\
&= S_{{N}-t} \theta \log (k_t) + R_t
\end{align*}
Hence, the lemma follows.
\end{proof}

\begin{lemma}\label{optkt}
The optimal trajectory of state $k_t^*[N]$ for the Bellman equation (\ref{Bellman}) is as follows:
\begin{align*}
\log k_t^*[N] = \sum_{i=1}^{t} \theta^{t-i} \log \left( \frac{S_{{N}-i}}{S_{{N}-i+1}} {A} \beta \theta \right) + \theta^{t}\log k_0
\end{align*}
\end{lemma}
\begin{proof}
We show this by induction.
By plugging (\ref{ctkt}) into (\ref{const}), we obtain:
\begin{align}
k_{t+1} = {A} (k_t)^\theta - c_t 
= \frac{S_{{N}-t} -1}{S_{{N}-t}}{A} (k_t)^\theta
= \frac{S_{{N}-t-1}}{S_{{N}-t}}{A}\beta\theta(k_t)^\theta
\label{ktt}
\end{align}
As we apply $t=0$ to the above (\ref{ktt}) and take the logarithm, 
\begin{align*}
\log k_1 
=\ln \left( \frac{S_{{N}-1}}{S_{N}} {A} \beta \theta \right) 
+ \theta \log k_0
\end{align*}
We know that the lemma is true for $t=1$.
Suppose that the lemma is true for $t$.
We then know by (\ref{ktt}) that:
\begin{align*}
\log k_{t+1}^*[N] 
&= \log \left(\frac{S_{{N}-t-1}}{S_{{N}-t}}{A} \beta \theta \right) + \theta \log k_t^*[N]  
\\
&=\log \left(\frac{S_{{N}-t-1}}{S_{{N}-t}}{A} \beta \theta \right)
+\theta \left( \sum_{i=1}^{t} \theta^{t-i} \log \left( \frac{S_{{N}-i}}{S_{{N}-i+1}} {A} \beta \theta \right) 
+ \theta^{t}\log k_0 \right)
\\
&=\sum_{i=1}^{t+1} \theta^{t+1-i} \log \left( \frac{S_{{N}-i}}{S_{{N}-i+1}} {A} \beta \theta \right) 
+ \theta^{t+1}\log k_0
\end{align*}
which indicates that the lemma is true for $t+1$.
Hence, the lemma follows.
\end{proof}

\begin{propo}\label{logoptct}
The optimal consumption trajectory $c_t^*[N]$ for the Bellman equation (\ref{Bellman}) is as follows:
\begin{align*}
\log c_t^*[N]
=\log \left( \frac{{A} }{S_{{N}-t}} \right) + \theta \left( \sum_{i=1}^{t} \theta^{t-i} \log \left( \frac{S_{{N}-i}}{S_{{N}-i+1}} {A} \beta \theta \right) + \theta^{t}\log k_0 \right)
\end{align*}
\end{propo}
\begin{proof}
This is obvious from Lemma \ref{optkt} and (\ref{ctkt}).
\end{proof}

\section{} \label{apd3}
Let us evaluate the following:
\begin{align*}
\lim_{\beta\to1} \frac{\beta- ((1-\beta){N} + 1)\beta^{N+1}}{(1-\beta)^2} 
&= \lim_{\beta\to1} \frac{1-(N+1)((1-\beta)N+1)\beta^N + N\beta^{N+1}}{-2(1-\beta)} 
\\
&= \lim_{\beta\to1} \frac{N(N+1)\left( 2\beta^N - ((1-\beta)N+1)\beta^{N-1} \right)}{2} 
\\
&= {N(N+1)}/{2}
\end{align*}
where we use L'H\^{o}pital's rule twice.
Then, we know that 
\begin{align*}
\lim_{\beta\to1} \frac{\beta- ((1-\beta){N} + 1)\beta^{N+1}}{(1-\beta)^2} \log(A\beta)
= \frac{N(N+1)}{2}\log(A) =0
\end{align*}

\section{\textmd{(Continuous-Time Extension and Optimal Stopping under CLU)}} \label{apd4}

To demonstrate the robustness of our finite-horizon result and to clarify its theoretical foundations, this appendix develops a continuous-time version of our model with a general Constant Relative Risk Aversion (CRRA) utility function.
It is important to delimit the scope of this appendix. Since we assume a linear technology and $\rho > 0$, this continuous-time formulation represents the exact analogue of the AK setting discussed in Section 3.1. Furthermore, we restrict our analysis to the parameter space $0 < \gamma < 1$. If $\gamma > 1$, the utility function is strictly negative globally, making immediate termination trivially optimal.

In standard optimal growth models, it is common to assume homothetic utility functions where the biological subsistence level approaches zero ($\nu \to 0$) \citep[e.g.,][]{CORDOBA}, implying that existence always yields strictly positive individual utility. However, under classical utilitarianism ($\alpha = 0$), maximizing welfare in a low-productivity economy over an infinite horizon mathematically forces consumption toward zero while extending the dynastic lifespan indefinitely ($N \to \infty$). This trajectory represents the dynamic analogue of the Repugnant Conclusion, where an infinite population living at the absolute margin of subsistence is ethically preferred to a smaller, prosperous one.

To circumvent this infinite accumulation of sub-optimal lives, we employ Critical-Level Utilitarianism (CLU). By introducing a strictly positive critical level ($\alpha > 0$), CLU establishes a social threshold ($\kappa$) for a life worth living. Crucially, this framework prevents the Repugnant Conclusion even when the underlying utility function is homothetic. Under CLU, if economic constraints force consumption below $\kappa$, the social evaluation of those generations becomes negative, despite their individual utility remaining technically positive.

To prevent the infinite accumulation of negative social value (a sub-critical existence), the optimal policy is to endogenously terminate the dynasty. This logic directly parallels the ethical proposition by \citet{Fleurbaey2014}, which posits that when welfare falls below a neutral threshold, non-existence is strictly preferred. Mathematically, this altruistic termination can be formulated as an irreversible optimal stopping problem, utilizing the two-stage optimal control framework developed by \citet{Tomiyama1985} and widely applied in macroeconomic regime-switching models \citep[e.g.,][]{Boucekkine2013}. Since the ethical value of non-existence is exactly zero under CLU, the value function for the post-termination stage is naturally set to zero ($V_2 = 0$, yielding the transversality condition $\mathcal{H}_1(T) = \mathcal{H}_2(T) = 0$).

In what follows, we solve this continuous-time optimal stopping problem to prove that the optimality of a finite horizon is a robust structural consequence of CLU, rather than an artefact of a specific functional form (such as the logarithmic utility).

\subsubsection*{The Continuous-Time Problem and Hamiltonian}
Consider a dynamic optimization problem with a free terminal time $T$:
\begin{align*}
    \max_{c_t, T} \quad & \int_{0}^{T} e^{-\rho t} U(c_t) dt \\
    \text{s.t.} \quad & \dot{k}_t = {F}(k_t) - c_t, \quad k_T = 0, \quad k_0 > 0 \text{ given} \nonumber
\end{align*}
where $\rho > 0$ is the discount rate. We employ a general CRRA utility function incorporating the CLU critical level $\alpha > 0$:
$
    U(c) = \frac{c^{1-\gamma} - 1}{1-\gamma} - \alpha
$
with the coefficient of relative risk aversion $\gamma = -c_t U''(c_t) / U'(c_t) \in (0, 1)$.

To solve this, we construct the present-value Hamiltonian $\mathcal{H}$:
\begin{align*}
    \mathcal{H}(c_t, k_t, \lambda_t, t) = e^{-\rho t} U(c_t) + \lambda_t ({F}(k_t) - c_t)
\end{align*}
where $\lambda_t$ is the costate variable representing the shadow price of capital.

\subsubsection*{First-Order Conditions and the Euler Equation}
The first-order necessary conditions for maximization are given by:
\begin{align}
    \frac{\partial \mathcal{H}}{\partial c_t} = 0 \quad &\implies \quad e^{-\rho t} U'(c_t) = \lambda_t \label{eq:foc_c} \\
    \dot{\lambda}_t = -\frac{\partial \mathcal{H}}{\partial k_t} \quad &\implies \quad \dot{\lambda}_t = - \lambda_t {F}'(k_t) \label{eq:foc_k}
\end{align}
Differentiating Eq. (\ref{eq:foc_c}) with respect to time yields:
\begin{align}
    \dot{\lambda}_t = -\rho e^{-\rho t} U'(c_t) + e^{-\rho t} U''(c_t) \dot{c}_t \label{eq:lambda_dot}
\end{align}
Equating (\ref{eq:lambda_dot}) with (\ref{eq:foc_k}) and rearranging provides the standard Euler equation:
\begin{align*}
    \frac{\dot{\lambda}_t}{\lambda_t} = -\rho + \frac{U''(c_t)}{U'(c_t)} \dot{c}_t = -\rho - \gamma \frac{\dot{c}_t}{c_t} = - {F}'(k_t) \quad \implies \quad \frac{\dot{c}_t}{c_t} = \frac{{F}'(k_t) - \rho}{\gamma}
\end{align*}

\subsubsection*{Explicit Solution for Linear Technology}
For analytical tractability, assume a linear production function ${F}(k_t) = r k_t$, where $r > 0$ is the constant marginal product of capital. The Euler equation simplifies to a constant growth rate:
\begin{align}
    \frac{\dot{c}_t}{c_t} = \frac{r - \rho}{\gamma} \quad \implies \quad c_t = C e^{\frac{r - \rho}{\gamma} t} \label{eq:c_path}
\end{align}
where $C$ is the initial consumption $c_0$. Substituting this into the capital accumulation equation gives a first-order linear differential equation:
\begin{align*}
    \dot{k}_t - r k_t = -C e^{\frac{r - \rho}{\gamma} t}
\end{align*}
Multiplying both sides by the integrating factor $e^{-rt}$ and integrating from $0$ to $T$ yields:
\begin{align*}
    \int_{0}^{T} \frac{d}{dt} \left( k_t e^{-rt} \right) dt = -C \int_{0}^{T} e^{\frac{(1-\gamma) r - \rho}{\gamma} t} dt
\end{align*}
Applying the terminal condition $k_T = 0$, we solve for the constant $C$:
\begin{align}
    -k_0 = -C \frac{\gamma}{(1-\gamma) r - \rho} \left( e^{\frac{(1-\gamma) r - \rho}{\gamma} T} - 1 \right) \quad \implies \quad C = \frac{\zeta k_0}{e^{\zeta T} - 1} \label{eq:C_val}
\end{align}
where we define $\zeta \equiv \frac{(1-\gamma) r - \rho}{\gamma}$. Consequently, the optimal consumption path for a given horizon $T$ is:
\begin{align}
    c_t = \frac{\zeta k_0}{e^{\zeta T} - 1} e^{\frac{r - \rho}{\gamma} t} \label{eq:ct_final}
\end{align}

\vspace{1em}
\noindent 
\textsl{Remark on the Infinite Horizon Paradox:}
\\
Equation (\ref{eq:C_val}) reveals a critical structural flaw of assuming an infinite planning horizon ($T \to \infty$) in this macroeconomic setting. If $(1-\gamma) r > \rho$ (i.e., $\zeta > 0$), taking the limit $T \to \infty$ forces the initial consumption $C$ to approach exactly zero. Under CLU, since $U(0) = -\alpha < 0$, enforcing a perpetual existence constraint would theoretically dictate a complete collapse of consumption, leading to an infinite accumulation of negative social value. This mathematical property underscores the necessity of endogenizing the terminal time.

\subsubsection*{Optimal Stopping and the Transversality Condition}
To determine the optimal longevity of the dynasty $T^*$, we consider the marginal change in the value function $V(T)$ with respect to an extension of the terminal time $dT$:
\begin{align*}
    dV(T) &= e^{-\rho T} U(c_T) dT + \lambda_T dk_T \nonumber \\
          &= e^{-\rho T} U(c_T) dT + \lambda_T \dot{k}_T dT \nonumber \\
          &= \left[ e^{-\rho T} U(c_T) + \lambda_T ({F}(k_T) - c_T) \right] dT
\end{align*}
Hence, $dV(T)/dT = \mathcal{H}(T)$. According to the principles of two-stage optimal control and optimal stopping, the Hamiltonian must evaluate to zero at the optimal stopping time $T^*$:
\begin{align*}
    \mathcal{H}(T^*) = e^{-\rho T^*} U(c_{T^*}) + \lambda_{T^*} ({F}(k_{T^*}) - c_{T^*}) = 0
\end{align*}
Using the terminal condition $k_{T^*} = 0$ (implying ${F}(k_{T^*}) = 0$) and substituting $\lambda_{T^*} = e^{-\rho T^*} U'(c_{T^*})$, this condition simplifies to:
\begin{align*}
    U(c_{T^*}) - U'(c_{T^*}) c_{T^*} = 0
\end{align*}
Applying the CRRA utility function $U(c) = \frac{c^{1-\gamma} -1}{1-\gamma} - \alpha$, the optimal terminal consumption level is derived as:
\begin{align}
    c_{T^*} = \left( \frac{(1-\gamma) \alpha + 1}{\gamma} \right)^{\frac{1}{1-\gamma}} \label{eq:cT_opt}
\end{align}
(Note that as $\gamma \to 1$, $c_{T^*} \to e^{1+\alpha}$, which is consistent with the logarithmic utility case).

Finally, equating the terminal consumption from the optimal trajectory (\ref{eq:ct_final}) evaluated at $t=T^*$ with the critical stopping condition (\ref{eq:cT_opt}), we obtain the implicit equation determining the optimal finite horizon $T^*$:
\begin{align*}
    \left( \frac{(1-\gamma) \alpha + 1}{\gamma} \right)^{\frac{1}{1-\gamma}} = \frac{\zeta k_0}{e^{\zeta T^*} - 1} e^{\frac{r - \rho}{\gamma} T^*}
\end{align*}
This result confirms that a finite, optimal planning horizon $T^*$ exists robustly under general CRRA preferences. The termination is mathematically driven by the social critical level $\alpha > 0$, avoiding the repugnant implications of the perpetual existence constraint.

\section*{Compliance with Ethical Standards}
The authors declare that they have no conflicts of interest.

\section*{Acknowledgements}
We would like to express our sincere gratitude to the Editor, Gr\'egory Ponthi\`ere, for his invaluable guidance and insightful comments that profoundly enriched the theoretical framework of this paper. We are also deeply grateful to the anonymous reviewers for their rigorous evaluation and highly constructive feedback. 
Any remaining errors are our own.

\bibliographystyle{spbasic_x}      
{\raggedright
\bibliography{bibfile}
}
\end{document}